\documentclass[aps,pra,twocolumn,superscriptaddress]{revtex4-2}

\usepackage{graphicx}  % needed for figures
\usepackage{epstopdf}

\usepackage{dcolumn}   % needed for some tables
\usepackage{bm}        % for math
\usepackage{amsmath}   % for math
\usepackage{amssymb}   % for math
\usepackage{maybemath}
\usepackage{xcolor}
\usepackage{enumitem}
\usepackage{graphicx}  % needed for figures
\DeclareUnicodeCharacter{2212}{-}

\newcommand{\bra}[1]{\langle #1|}
\newcommand{\ket}[1]{|#1\rangle}

\newcommand{\rarr}[0]{\rightarrow}

\newcommand{\eps}[0]{\varepsilon}
\newcommand{\pphi}[0]{\varphi}
\newcommand{\up}[0]{\uparrow}
\newcommand{\dn}[0]{\downarrow}
\newcommand{\dw}[0]{\downarrow}
\newcommand{\ho}[0]{\mathrm{ho}}
\newcommand{\lu}[0]{\mathrm{lu}}

\newcommand{\s}[0]{\sigma}

\newcommand{\KS}[0]{\mathrm{KS}}
\newcommand{\LSDA}[0]{{L\!S\!D\!A}}
\newcommand{\PBE}[0]{{P\!B\!E}}
\newcommand{\DSCF}[0]{{\Delta SC\!F}}

\newcommand{\Hxc}[0]{\mathrm{Hxc}}

\newcommand{\leqs}[0]{\leqslant}

\newcommand{\rr}[0]{\mathbf{r}}

\newcommand{\ens}{\textrm{ens}} 
\newcommand{\IP}{I\!P} 
\newcommand{\EA}{E\!A} 

\begin{document}

\title{Ionization Potentials and Fundamental Gaps in Atomic Systems from the Ensemble-DFT Approach}

\author{Sharon Lavie}
%\email{Sharon.Lavie@mail.huji.ac.il}
\affiliation{Fritz Haber Center for Molecular Dynamics and Institute of Chemistry, The Hebrew University of Jerusalem, 9091401 Jerusalem, Israel}

\author{Yuli Goshen}
\affiliation{Fritz Haber Center for Molecular Dynamics and Institute of Chemistry, The Hebrew University of Jerusalem, 9091401 Jerusalem, Israel}

\author{Eli Kraisler}
\email{Author to whom correspondence should be addressed: eli.kraisler@mail.huji.ac.il}
\affiliation{Fritz Haber Center for Molecular Dynamics and Institute of Chemistry, The Hebrew University of Jerusalem, 9091401 Jerusalem, Israel}

\date{\today}

\begin{abstract}
Calculations in Kohn-Sham density functional theory crucially rely on high-quality approximations for the exchange-correlation (xc) functional. Standard local and semi-local approximations fail to predict the ionization potential (IP) and the fundamental gap, departing from the Kohn-Sham orbital energies, due to the deviation of the total energy from piecewise-linearity and the absence of the derivative discontinuity. 
The ensemble generalization procedure introduced in Phys.\ Rev.\ Lett.\ 110, 126403 (2013) restores, to a large extent, these features in any approximate xc functional and improves its ability to predict the IP and the fundamental gap with negligible additional computational effort. In this work we perform an extensive study of atoms and first ions across the Periodic Table, generalizing the local spin-density and the Perdew-Burke-Ernzerhof approximations. By applying the ensemble generalization to a variety of systems, with $s$-, $p$- and $d$-character, we assess the accuracy of the method and identify important trends. In particular, we find that the accuracy of our approach heavily depends on the character of the frontier orbitals: when $d$-orbitals are involved, the performance is far less accurate.  Possible sources of error are discussed and ways for further improvement are outlined.

\end{abstract}

\maketitle 

\section{Introduction}

Over the last decades, density functional theory (DFT) became the leading theoretical framework for quantum simulations of atoms, molecules, and solids~\cite{DG,Burke12,Marzari21}.  DFT offers an in-principle exact and numerically effective scheme to calculate ground-state properties of many-electron systems. It is based on the fundamental findings of Hohenberg and Kohn~\cite{HK64} and Kohn and Sham (KS)~\cite{KS65}, which allow to reformulate the problem of a many-electron interacting system into a set of one-electron Schrödinger-like equations that can be solved self-consistently. KS equations include in principle all the many-body effects through a multiplicative exchange-correlation (xc) potential. This potential is unknown, and constitutes the central approximation in DFT. Understanding of many central features of electronic structure, such as bonding, ionization, dissociation, magnetism, etc.\ requires an accurate description of exchange and correlation. 

Atoms and ions are a convenient and promising class of systems to analyze the validity and accuracy of xc approximations. On the one hand, atoms are the simplest possible units in chemistry that are relatively easy to assess numerically, which is why DFT studies on atoms have a long history~\cite{TongSham66,MJW,Kotochigova97,LeeMartin97,ErnzerhofScuseria99,KraislerMakovKelson10,Klupfel11,Argaman13,Lehtola20}. On the other hand, by considering all atoms of the Periodic Table, one obtains a rich class of systems, with $s$-, $p$-, $d$- and $f$-electrons. In addition, the experimental data on atomic systems are abundant and are generally very precise~\cite{HandChemPhys95}.

%While the simplest approximation for the xc energy functional is the local spin density approximation (LSDA)~\cite{ParrYang84} of homogeneous electron gas, which has been widely used in DFT structure calculations.

Common xc energy functionals, e.g., the local spin-density approximation (LSDA)~\cite{PW92} and the semi-local Perdew-Burke-Ernzerhof (PBE) generalized gradient approximation (GGA)~\cite{PBE96} proved to be rather accurate in total energy calculations~\cite{Kotochigova97, KraislerMakovArgamanKelson09, KraislerMakovKelson10, Argaman13, Lehtola20}: %\blue{\cite{Vydrov07, Talman76}}
The ionization potentials (IP) of atoms and ions can be obtained from total energy differences, $\IP = E(N_0-1)-E(N_0)$, within a few percent accuracy. 
However, as one attempts to calculate the IP from the highest occupied (ho) orbital energy, $\eps_\ho$, %EK: relying on the IP theorem in DFT [REFs], 
within the same xc approximations, the result is typically $40-50\%$ too low~\cite{GunnJones80,TongChu97,Grabo97,Chan99,AllenTozer02,Teale08,SalzBaer09,Klupfel11}. Therefore, correct prediction of the homo energy~\footnote{We refer to $\eps_\ho$ as the \emph{ho} energy or as the \emph{homo} energy.} is a serious challenge for many xc approximations. Another, closely related, challenge is to obtain the fundamental gap, $E_g$, which is the difference between the IP and the electron affinity (EA), from a single DFT calculation. 
Unlike the IP, the EA has no direct analog in the KS system; it does not equal the negative of the lowest unoccupied (lu) orbital energy and hence the fundamental gap does not equal the KS gap, $E_g^\KS = \eps_\lu - \eps_\ho$, \emph{even for the exact xc functional.} Approximating $E_g$ by $E_g^\KS$ not only results in an underestimation of the gap, but sometimes leads to qualitatively wrong results, e.g.\ %EK: wrongly 
predicting an insulating system to be metallic (gapless). This article focuses on improving the prediction for the IP and $E_g$ in atoms and ions, departing from the KS orbital energies $\eps_{i\s}$, using the ensemble approach~\cite{KraislerKronik13}, as explained below.

\emph{Piecewise-linearity, IP theorem and the discontinuity.}
Accurate prediction of both the IP and $E_g$ with an approximate xc functional closely relates to the expected piecewise-linear (PWL) behavior of the total energy versus number of electrons, $E(N)$~\cite{PPLB82,Yang00}. For the exact xc functional, when the number of electrons, $N$, varies, and can take on both integer and fractional values, the energy $E(N)$ changes linearly for any fractional $N$. In particular, if the neutral system has $N_0$ electrons, then for $N \in [N_0-1,N_0]$ the slope of $E(N)$ is $E(N_0) - E(N_0-1) = -\IP$, whereas for  $N \in [N_0,N_0+1]$ the slope equals $E(N_0+1) - E(N_0) = -\EA$. %Thus, the slope of $E(N)$ equals the corresponding energy difference and changes as $N$ crosses an integer. 
In general, the slope of $E(N)$ changes as $N$ crosses an integer.
The discontinuity in the slope %of $E(N)$ 
at $N=N_0$ equals the fundamental gap, $E_g = \IP - \EA$.
Similarly to the total energy, any expectation value of the ground state is also piecewise linear, e.g.\ the external potential energy, $\int v_\textrm{ext}(\rr) n(\rr) d^3r$. Furthermore, Ref.~\cite{GouldToulouse14} has shown that for selected atomic systems the non-interacting kinetic energy, although \emph{not} an expectation value of the interacting wavefunction, is also approximately piecewise linear.

An important consequence %manifestation 
of piecewise linearity is the relation between the IP, as the slope of $E(N)$, and $\eps_\ho$, which can be established via Janak's theorem~\cite{Janak78}, or via the density decay rate (see, e.g., \cite{LevyPerdewSahni84,AlmbladhVonBarth85,GoriGiorgi16,GoriGiorgi18,HodgsonKraisler17,KraislerHodgson20,Kraisler20} and references therein). If piecewise linearity is maintained, then 
\begin{equation}
\IP = -\eps_\ho(N_0^-),    
\end{equation}
a result known as the IP theorem of DFT~\cite{PPLB82,AlmbladhVonBarth85, LevyPerdewSahni84, PerdewLevy97, Harbola99}. 
We note that the IP theorem has recently been extended to ensembles of excited states (see Ref.~\cite{Gould22} and their Supporting Information).

In contrast to the IP, the EA does not equal the (negative of) $\eps_\lu(N_0^-)$. Instead, $\EA = -\eps_\lu(N_0^-) - \Delta$, where $\Delta$ is the abrupt spatial ``jump'' experienced by the KS potential as $N$ surpasses an integer~\cite{PPLB82,PerdewLevy83,Godby87,Godby88,GouldToulouse14,Morrison15}. It is termed the derivative discontinuity (DD). As a result, the fundamental gap is expressed in terms of KS-DFT as
\begin{equation} \label{eq:Eg}
E_g = E^\KS_g + \Delta.
\end{equation}

\emph{Reproducing the DD in approximations.}
Despite the importance of piecewise linearity, commonly used functional classes, such as the local spin-density approximation (LSDA), the  generalized gradient approximations (GGAs), and also most hybrid approximations, grossly disobey the PWL condition. Instead, a typically convex $E(N)$ curve is obtained~\cite{SteinKronikBaer_curvatures12,PerdewLevy97,KraislerKronik13,Atalla16}, and correspondingly, $\eps_\ho$ underestimates the IP~\cite{AllenTozer02,Teale08,KK08}. This happens not only in atoms, but also in molecular systems.  Furthermore, since for many functionals the xc potential varies smoothly with $N$ and the DD is absent, the fundamental gap is also strongly underestimated.

Various methods within DFT have been suggested to correct these drawbacks. Usage of new types of GGAs can yield improved properties of the xc potential, positively affecting the KS spectrum~\cite{ArmientoKummel13, Vlcek15, Tran15}, but at the cost of being less accurate for total energies. Meta-GGAs, where the xc energy density at each point depends not only on the electron density and its gradient, but on additional quantities, particularly on kinetic energy densities~\cite{TPSS03,Sun_SCAN_PRL15}, produce more realistic gaps~\cite{Yang16,Perdew_gap_17,AschebrockKummel19}, particularly within the generalized KS (GKS) scheme, where the multiplicative KS potential is replaced by a non-local operator \cite{Seidl96_GKS}.  
Functionals that are explicitly orbital-dependent and inherently carry a derivative discontinuity, have been extensively explored. Calculations with this class of functionals can be done either within the GKS scheme, or remaining within the KS scheme, using the optimized effective potential procedure~\cite{SharpHorton53, Talman76, Grabo97, KK08}, which can be computationally expensive.
Global hybrid functionals~\cite{AdamoBarone99, PerErnBurke96, ErnzerhofScuseria99}, which linearly combine (semi-)local xc energy components and Fock exchange, can in principle be fitted such that they produce a piecewise-linear energy curve and an improved value for the homo, but this happens when using $\approx 75$\% of Fock exchange~\cite{Sai11,Atalla13}, which in most cases significantly compromises the performance of the functional for other, total-energy-related 
quantities~\cite{Kronik_JCTC_review12}.
An alternative is local hybrid functionals, where the exact exchange and a semi-local exchange and correlation are mixed locally~\cite{Cruz98,Jaramillo03,SchmidtKraisler_ISO_14,SchmidtPCCP14,MaierArbuzKaupp19}, allowing e.g., to treat those regions in space where there is essentially one electron with Fock exchange and regions with a slowly varying density with a semi-local functional. For example for the ISOcc local hybrid~\cite{SchmidtKraisler_ISO_14}, when fitted to reproduce atomization energies of diatomics, the deviation of $-\eps_\ho$ from experiment is about 25\%. Fitting to optimize the prediction of the IP significantly reduces the accuracy of total-energy-related properties, similar to global hybrids. 
Another alternative is range-separated hybrids that can be optimally tuned, per system, to closely reproduce the IP theorem and introduce the DD~\cite{Eisenberg09,Stein10,BaerLivSalz10,Kronik_JCTC_review12,Refaely13}.
There also exist a number of schemes which modify a given xc approximation by enforcing a set of constraints, for instance the scaling correction which aims to maintain piecewise linearity \cite{Zheng11, Zheng13, Li18,Mahler22}, DFT+U methods~\cite{Coco05,KulikCocoMarzari06,LanyZunger09} and self-interaction corrections~\cite{PZ81,KronikKummel20}, which seek to eliminate one-electron self-interaction, yielding a significant improvement
in the interpretation of KS eigenvalues~\cite{Korzdorfer09}. Yet, their performance for ground-state
energetics is debatable~\cite{Vydrov04, Vydrov06a, Hofmann12a, Klupfel12}. Another such approach is Koopmans-compliant functionals, which reduce self-interaction errors by requiring that changing the occupation of any KS orbital will not affect its energy eigenvalue~\cite{Dabo10,Dabo14,Borghi14,ColonnaNguyen19}.

\emph{Ensemble Generalization~\cite{KraislerKronik13}. }
 A promising approach that closely reconstructs the PWL condition from first principles, on which we focus in this paper, %and thus improves the prediction of the IP via the homo, and reintroduces a DD, 
 termed the ensemble generalization, has been suggested in Ref.~\cite{KraislerKronik13}. This approach relies on the fact that both the interacting and the KS systems have to be described by ensembles, for any fractional $N$. Consequently, the Hartree and xc energy functionals have to be appropriately generalized. The approach comes completely from first principles and does not introduce any empiricism. It can be applied to any underlying xc approximation.
 
 The ensemble generalization approach of Ref.~\cite{KraislerKronik13} was able to closely reconstruct the PWL behavior of the total energy for various approximate functionals: local and semi-local approximations as well as global and local hybrids. As a result, improvement in the correspondence between $-\eps_\ho$ and the experimental IP was achieved and reported in Ref.~\cite{KraislerSchmidt15} for a small set of diatomic molecules and constituent atoms. However, this approach has not been applied on a larger scale so far.
 
 In addition, within the ensemble generalization approach an analytic expression for the DD has been derived~\cite{KraislerKronik14}. This allows us to calculate the fundamental gap relying on KS quantities of the neutral system only, thus avoiding calculations of the total energy of three systems:  the neutral  system,  the  cation,  and  the  anion, which is particularly useful when one of the ionic calculations is numerically challenging. The ensemble-DD correction suggested in Ref.~\cite{KraislerKronik14} was not used so far.

In this article we perform a comprehensive study of atoms and their first ions across the Periodic Table ($1 \leqs Z \leqs 88$), applying the ensemble generalization~\cite{KraislerKronik13, KraislerKronik14}. We focus on two main properties: the IP and the fundamental gap, $E_g$. We analyze in detail the $s$-, $p$- and $d$-blocks of the Periodic Table and connect the accuracy of our results to the positions of the systems in the  Table. Furthermore, we discuss the remaining discrepancy with respect to experiment and outline possible future improvements in predicting ionization potentials and fundamental gaps in atomic systems.

The article is organized as follows. Section~\ref{sec:Theoretical Background} provides the relevant theoretical background on the ensemble approach. In Sec.~\ref{sec:NumericalDetails} the numerical details for the calculations are given. In Sec.~\ref{sec:Results} we present our numerical findings for atoms and ions, focusing on the IP in Secs.~\ref{sec:Results:IP} and~\ref{sec:Results:IP_d} and on the fundamental gap in Secs.~\ref{sec:Results:Eg_ions} and~\ref{sec:Results:Eg_neutral}. Finally, in Sec.~\ref{sec:Summary} we summarize %EK: the present 
our work.

\section{Theoretical Background} \label{sec:Theoretical Background}
%a.	Ensemble state
%b.	Generalization of the Hxc functional
%c.	KS potential 
%d.	Hxc potential
%e.	DD
%f.	An approximation for the DD-ensemble treatment  

The ensemble generalization approach used in this work was proposed by Kraisler and Kronik in Ref.~\cite{KraislerKronik13} and further elaborated in Refs.~\cite{KraislerKronik14,KraislerSchmidt15,KraislerKronik15}. It is briefly summarized below, highlighting the findings that are relevant in the present context.

\emph{Interacting many-electron systems of fractional $N$.}
The ensemble generalization  formally analyzes the KS system with a varying, possibly non-integer number of electrons, $N$,  subsequently taking the limit of an integer $N$. An open many-electron system that can exchange electrons with its environment, and hence can possess on average a fractional number of electrons, \emph{cannot} be described by a pure quantum state, but rather by an \emph{ensemble} of states~\cite{PPLB82}. In the following, we focus on Coulomb systems at zero temperature and consider the case of a system with $N_0-1+\alpha$ electrons, where $\alpha \in [0,1]$. The case $N=N_0$ (i.e., $\alpha=1$) corresponds to the electrically neutral system and $N=N_0-1$ (i.e., $\alpha=0$) corresponds to the cation. Using the convexity conjecture~\cite{DG,Lieb,Cohen12,PPLB82}, which states that $\IP>\EA$ for any many-electron system, with any number of electrons, and further assuming that the ground states of the $N_0$- and $(N_0-1)$-electron systems are not degenerate, it has been shown~\cite{PPLB82} that the ground-state ensemble consists of two states: $\hat{\Lambda} = (1-\alpha) \ket{\Psi_{N_0-1}}\bra{\Psi_{N_0-1}} + \alpha \ket{\Psi_{N_0}}\bra{\Psi_{N_0}}$. Here $\ket{\Psi_{N_0-1}}$ and $\ket{\Psi_{N_0}}$ are the pure ground states of the cation and the neutral, respectively%We assume that both these ground states are not degenerate 
~\footnote{In the presence of degeneracy, different choices of $\ket{\Psi_{N_0-1}}$ and $\ket{\Psi_{N_0}}$ will result in the same total energy, but different ensemble densities, and hence different KS potentials. In this work, however, we confine ourselves to the two-state ensemble only.
%$\ket{\Psi_{N_0-1}},\ket{\Psi_{N_0}}$ are simply any two ground states with $N_0-1$ and $N_0$ electrons, respectively. Different choices of $\ket{\Psi_{N_0-1}},\ket{\Psi_{N_0}}$ will result in the same total energy, but different densities, and hence different KS potentials.}
}. 
This form of $\hat{\Lambda}$ was used to derive the piecewise-linearity requirements of the energy and the density~\cite{PPLB82}.

\emph{Kohn-Sham system of fractional $N$.}
In the KS system the number of electrons is equal to that of the real system. Therefore, the number of KS electrons can be fractional, as well. Then, inevitably, the KS system must also be described by an ensemble: $\hat{\Lambda}^\KS = (1-\alpha) \ket{\Phi_{N_0-1}^{(\alpha)}}\bra{\Phi_{N_0-1}^{(\alpha)}} + \alpha \ket{\Phi_{N_0}^{(\alpha)}}\bra{\Phi_{N_0}^{(\alpha)}}$, where $\ket{\Phi_{N_0-1}^{(\alpha)}}$ and $\ket{\Phi_{N_0}^{(\alpha)}}$ %EK: adding kets
are ground states with $N_0-1$ and $N_0$ electrons, respectively. These two ground states belong to the same KS system, defined by the potential $v_\KS^{(\alpha)}(\rr)$. Therefore, the wavefunctions $\Phi_{N_0-1}^{(\alpha)}$ and $\Phi_{N_0}^{(\alpha)}$ are Slater determinants consisting of the $N_0-1$ and $N_0$ lowest-lying KS orbitals, $\pphi_i^{(\alpha)}(\rr)$, respectively. The superscript $(\alpha)$ emphasizes that the KS orbitals (and the KS potential and all quantities that are created from them) are $\alpha$-dependent.
Fully acknowledging the fact that the KS system is in an ensemble state allows us to generalize any approximate Hartree-xc functional for ensembles, as follows~\cite{KraislerKronik13}:
\begin{equation} \label{eq:energy functional generalization}
E_{e-\Hxc}[n^{(\alpha)}] = (1-\alpha)E_{\Hxc}[{\rho^{(\alpha)}_{-1}}]  + \alpha E_{\Hxc}[{\rho^{(\alpha)}_{0}}].
\end{equation}
Here $E_{\Hxc}$ is the usual, pure-state Hartree-xc functional, the index $e-\Hxc$ indicates the functional is ensemble-generalized, the quantities $\rho^{(\alpha)}_{-1} (\rr)= \sum_{i=1}^{N_0-1} |\pphi_i^{(\alpha)}(\rr)|^2$ and $\rho^{(\alpha)}_{0}(\rr) = \sum_{i=1}^{N_0} |\pphi_i^{(\alpha)}(\rr)|^2$ are two auxiliary densities formed by summing up the first $N_0-1$ and $N_0$ squared orbitals, and the density $n^{(\alpha)}(\rr)$ is the ensemble many-electron density, which can be expressed as $n^{(\alpha)}(\rr) = (1-\alpha) \rho^{(\alpha)}_{-1}(\rr)  + \alpha \rho^{(\alpha)}_{0}(\rr)$. Equation~\ref{eq:energy functional generalization} is exact for the Hartree and the exact exchange components, and is approximate for the correlation~\cite{Goerling15,Kraisler_PhD}. The term that is neglected when enforcing Eq.~(\ref{eq:energy functional generalization}) on the correlation is related to the density-driven correlations derived in Ref.~\cite{GouldPittalis19}.

Employment of the ensemble generalization (Eq.~(\ref{eq:energy functional generalization})) showed %EK: great
improvement in satisfying the PWL criterion for the energy~\cite{KraislerKronik13,KraislerKronik15}. The obtained slopes of $E(N)$ are much closer to straight lines. But if the slope of $E(N)$ changes, the homo energy must change as well, according to Janak’s theorem~\cite{Janak78}. The change in $\eps_\ho$ comes from the generalization of the Hartree-xc potential. In the limit of $\alpha \rarr 1^-$ (neutral system), the ensemble-generalized Hxc potential, $v_{e-\Hxc}[n](\rr)$, is expressed as
\begin{equation} \label{eq:Hxc potential generalization}
v_{e-\Hxc}[n](\rr) = v_{\Hxc}[n](\rr)  +v_{0}[n],
\end{equation}
where $v_{\Hxc}[n](\rr)$ is the original Hxc potential that stems from $E_{\Hxc}$, and
\begin{equation} \label{eq:spatially uniform term}
\begin{split}
v_{0} [n] & = E_{\Hxc}[n] - E_{\Hxc} [n-{|\pphi _\ho|}^{2}]  \\ 
& - \int |\pphi_\ho(\rr)|^{2} \,\, v_{\Hxc}[n](\rr) \,\, d^{3}r.
\end{split}
\end{equation}
%In words, the ensemble-generalized potential consists of two terms: the original, pure-state, $\rr$-dependent Hxc potential, $v_{Hxc}[n](\rr)$, and a spatially uniform term, $v_{0}$. %Notably, $v_{e-\Hxc}[n](\rr)$ does not decay to 0 at $r \rarr \infty$, but rather approaches a finite value.
Notably, in the integer-$N$ limit, the KS orbitals, the electron density, and the total energy will not be affected by the ensemble generalization: they will remain the same as those attained by the original Hxc functional. In contrast, the KS orbital energies will change -- they will be shifted by $v_0$: $\tilde{\eps}_i = \eps_i + v_0$. In particular, the new homo energies, $\tilde{\eps}_\ho$ are expected to more closely satisfy the IP theorem of DFT, and by that, the ionization potential can be obtained from the ho energy level with greater accuracy~\cite{KraislerSchmidt15}. 

The ensemble generalization also allows us to obtain an improved expression for the fundamental gap of the system. This is done via an analytic derivation for the DD~\cite{KraislerKronik14} with its subsequent use in Eq.~(\ref{eq:Eg}). The derivation relies on analyzing the behavior of the ensemble-generalized KS potential $v_{e-\Hxc}[n](\rr)$ around $N=N_0$. Namely, one extends the treatment described above also to $N \in (N_0,N_0+1]$ and then obtains the KS potential in the limit $ N \rarr N_0^+$ and $N \rarr N_0^-$. From the difference of these two limits one deduces the ensemble DD, $\Delta_\ens$~\cite{KraislerKronik14}: 
\begin{equation} \label{eq:delta ens}
\begin{split}
\Delta_\ens & = E_\Hxc[n+{|\pphi_{\lu}|}^{2}] - 2E_\Hxc[n] + E_\Hxc[n-|\pphi_\ho|^2] \\
&+ \int \left( |\pphi_\ho (\rr)|^2 - |\pphi_{\lu}(\rr)|^{2} \right) v_\Hxc[n](\rr) \,\, d^{3}r.
\end{split}
\end{equation}
Hence, the fundamental gap can be calculated as $E_g^\ens = E_g^\KS + \Delta_\ens$.
From Eq.~(\ref{eq:delta ens}) it follows that \emph{all} xc functionals, including the simplest LDA, do have a DD, which is revealed by the ensemble generalization. $\Delta_\ens$ can be calculated, with negligible numerical effort, from quantities that are available in any routine KS-DFT calculation: the ground-state density $n$, the corresponding Hxc potential $v_\Hxc[n](\rr)$ and the orbitals $\pphi_\ho (\rr)$ and $\pphi_{\lu}(\rr)$. Incorporating $\Delta_\ens$ in Eq.~(\ref{eq:Eg}) opens the gap and is expected to yield results that are much closer to experiment. 

\emph{Extension to the spin-dependent case.}
The results above have been presented in a spin-independent form, for simplicity. In practice, atomic systems usually have a nonzero spin, which needs to be taken into account. %EK:and therefore we treat them with spin-DFT. 
In the spin-dependent case, there exist two potential shifts, $v_0^\s$ (where $\s = \up, \dn$) for both spin channels, which use the corresponding spin-specific homo orbitals, $\pphi_\ho^\s(\rr)$. The global ho level is the highest between the two $\s$-ho levels: $\eps_\ho = \max_\s \left( \eps_\ho^\s \right)$, and similarly $\tilde{\eps}_\ho = \max_\s \left( \tilde{\eps}_\ho^\s \right)$.  Therefore, for calculating the IP with our approach, we ensemble-correct both homo levels and choose the highest among them. In most cases, the choice of the global homo does not change after the ensemble correction, apart from 7 atoms, as we detail in Sec.~\ref{sec:Results:IP_d}. 
We use similar considerations also when calculating the fundamental gap via
\begin{align} \label{eq:Delta_ens_tau_sigma}
\Delta_\ens^{\tau \s} &=  E_\Hxc[n+{|\pphi^\tau_\lu|}^{2}] - 2E_\Hxc[n] + E_\Hxc[n-|\pphi^\s_\ho|^2] \nonumber \\
&+ \int |\pphi^\s_\ho (\rr)|^2 v_\Hxc^\s (\rr) \,\, d^{3}r - \int  |\pphi^\tau_\lu (\rr)|^2  v_\Hxc^\tau (\rr) \,\, d^{3}r,
\end{align}
where $\s$ is the spin of the homo and $\tau$ -- of the lumo. Out of the four possible choices we favor the one that yields the minimal ensemble-corrected gap. Further details are given in Sec.~\ref{sec:Results:Eg_ions}.

%To summarize, in this work we obtain the IP and the fundamental gap using the ensemble generalization. For the IP, we shift the approximate KS homo energy by $v_0$ given in Eq.~(\ref{eq:spatially uniform term}). The fundamental gap is calculated using Eq.~(\ref{eq:Eg}), with the ensemble-DD from Eq.~(\ref{eq:delta ens}).

\section{Numerical Details} \label{sec:NumericalDetails}

All calculations were performed using the program package \texttt{ORCHID}~\cite{KraislerMakovKelson10}, version 6.1.1, which allows for electronic structure calculations of single atoms and ions, within a spherical approximation. %EK: logarithmic, not exp.
The effective Schr\"odinger equations are solved using a two-side shooting method on a discrete logarithmic grid. 
To achieve convergence of the total energy to $2 \cdot 10^{-6}$ Hartree and of the homo and lumo orbital energies, before and after the ensemble correction, to $10^{-5}$ Hartree, we used a grid with $N = 16000$ points, on the interval $[r_\textrm{min}, r_\textrm{max}]$, where $r_\textrm{min} = e^{-13}/Z$ Bohr,  with $Z$ being the atomic number, and $r_\textrm{max} = 35$ Bohr, for both LSDA and PBE-GGA. 

For systems where either $\eps_{\lu}^\up$ or $\eps_{\lu}^\dn$ was found to be greater than $-0.05$ Hartree, the original $r_\textrm{max}$ was too small to meet the convergence criteria, since $\pphi_\lu^\sigma(r) \sim e^{-\sqrt{-2\eps_\lu^\sigma}r}$  decays relatively slowly~\cite{LevyPerdewSahni84}. Therefore, for such systems $r_\textrm{max}$ was increased to $90$ Bohr, and $N$ to 17000 points.

Dealing with spin-dependent systems, we did not rely on experimental knowledge of the spin of the atoms and ions. For each system, we performed a series of calculations varying the ($z$-projection of the) total spin $S$, by changing the number of electrons in each spin channel, $N_\up$ and $N_\dw$, and finding the spin for which the total energy is minimal. For this sake, we started with the lowest possible value for $S$ (0 or 1/2, depending on the system), and increased it until a minimum in the total energy $E(S)$ was obtained (at $S_0$). Then, we performed calculations for at least one additional value, $S_0+1$, to "bracket" the minimum point. We had to assume, though, that no other minima in $E(S)$ are expected, for higher $S$. This assumption is in agreement with experimental data \cite{XD_Yang16}. Here and below we refer only to non-negative values of $S$, bearing in mind that in absence of a magnetic field $E(S)=E(-S)$.

\section{Results} \label{sec:Results}

%In this study, we aim to accurately obtain the IP and the fundamental gap from the KS eigenvalues, within a single DFT calculation. To this end, we apply the ensemble generalization  method~\cite{KraislerKronik13,KraislerKronik14} and obtain the IP as the sum of the highest occupied orbital energy, $\eps_\ho$, and the potential shift, $v_0$, and the fundamental gap as the sum of the KS gap and the ensemble derivative discontinuity, $\Delta_\ens$. 

In this section we present calculations of the IP for atoms with $1 \leqs Z \leqs 88$ and calculations of the fundamental gap for atoms and their first ions in this range. The raw data is available in the Supplemental Material. We consider two exchange-correlation energy functionals: the LSDA, in the  Perdew-Wang parametrization~\cite{PW92} and the GGA by Perdew, Burke and Ernzerhof (PBE)~\cite{PBE96}. We exclude from our study those atoms and ions whose KS ground state is not  pure-state-$v$-representable. In these systems, the $d$- (or $f$-) and $s$-levels of the KS system are degenerate and fractionally occupied (see, e.g.,~\cite{KraislerMakovKelson10} for details).  The excluded systems contain the well-known cases of Fe and Co, as well as Sc$^+$, Ti$^+$, Zr$^+$, Ba$^+$, Hf$^+$, and all the lanthanides, except Eu, Yb and Lu. Treatment of systems with fractional occupations is not covered by the approach presented above, because it assumes that both the neutral's and the ion's ground states are pure states, i.e. have integer occupations~\cite{KraislerKronik13}. Treatment of the systems we excluded above requires a further generalization of the method and is beyond of the scope of this work.

\subsection{Ionization potentials of neutral atoms with $1 \leqs Z \leqs 88$} \label{sec:Results:IP}
 
\begin{figure*}
  \centering
  \includegraphics[width=1.0\textwidth]{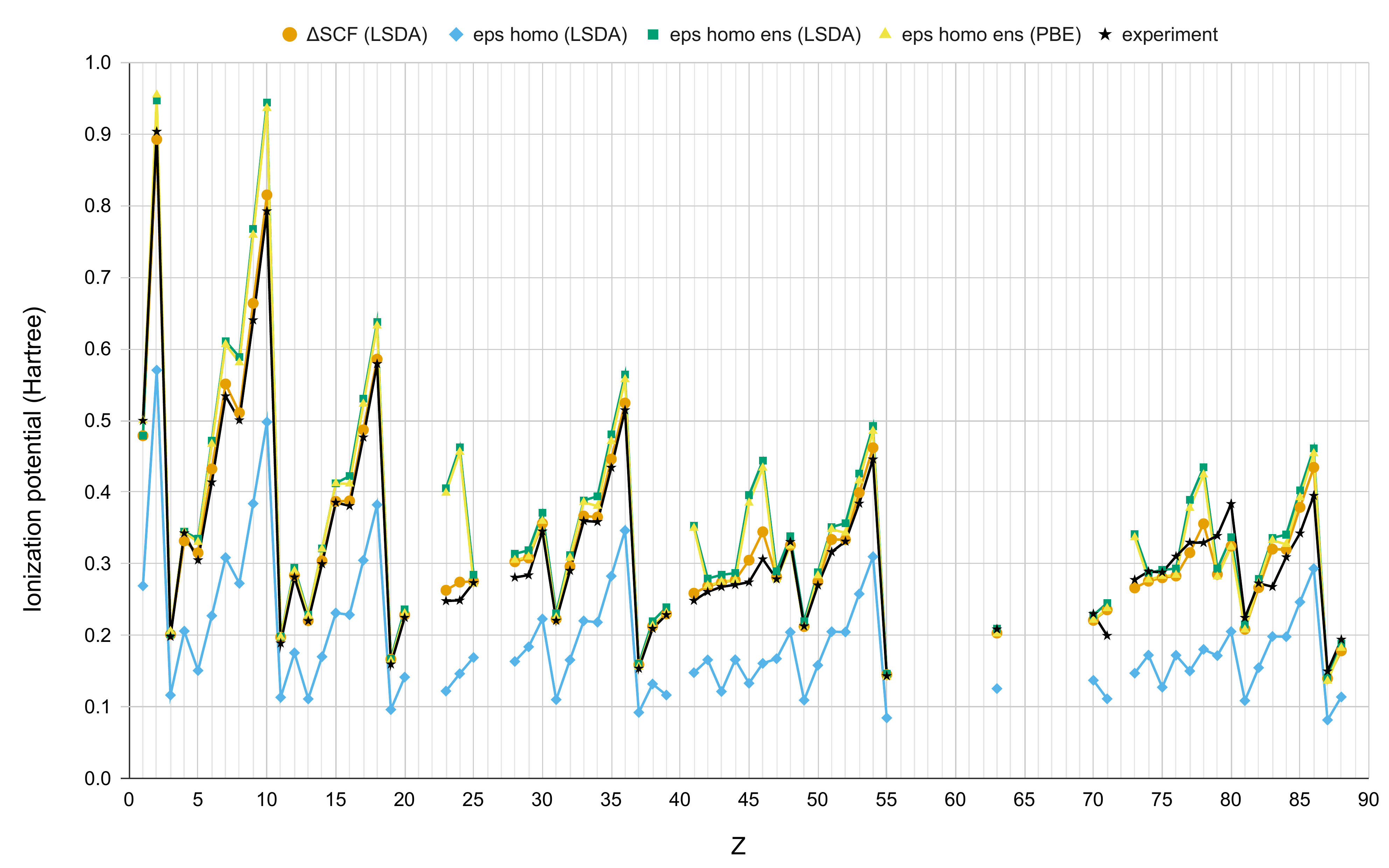}
\caption{The ionization potentials of neutral atoms, as a function of the atomic number $Z$, obtained from the homo energy levels, for the LSDA, before (light blue rhombuses), and after (green squares) the ensemble correction, and for the PBE after the ensemble correction (yellow triangles). Experimental values (black stars) and LSDA  $\DSCF$ results (orange circles) are shown for comparison.}
\label{fig:IP Vs.Z}
\end{figure*}

\begin{figure*}
  \centering
  \includegraphics[width=1.0\textwidth]{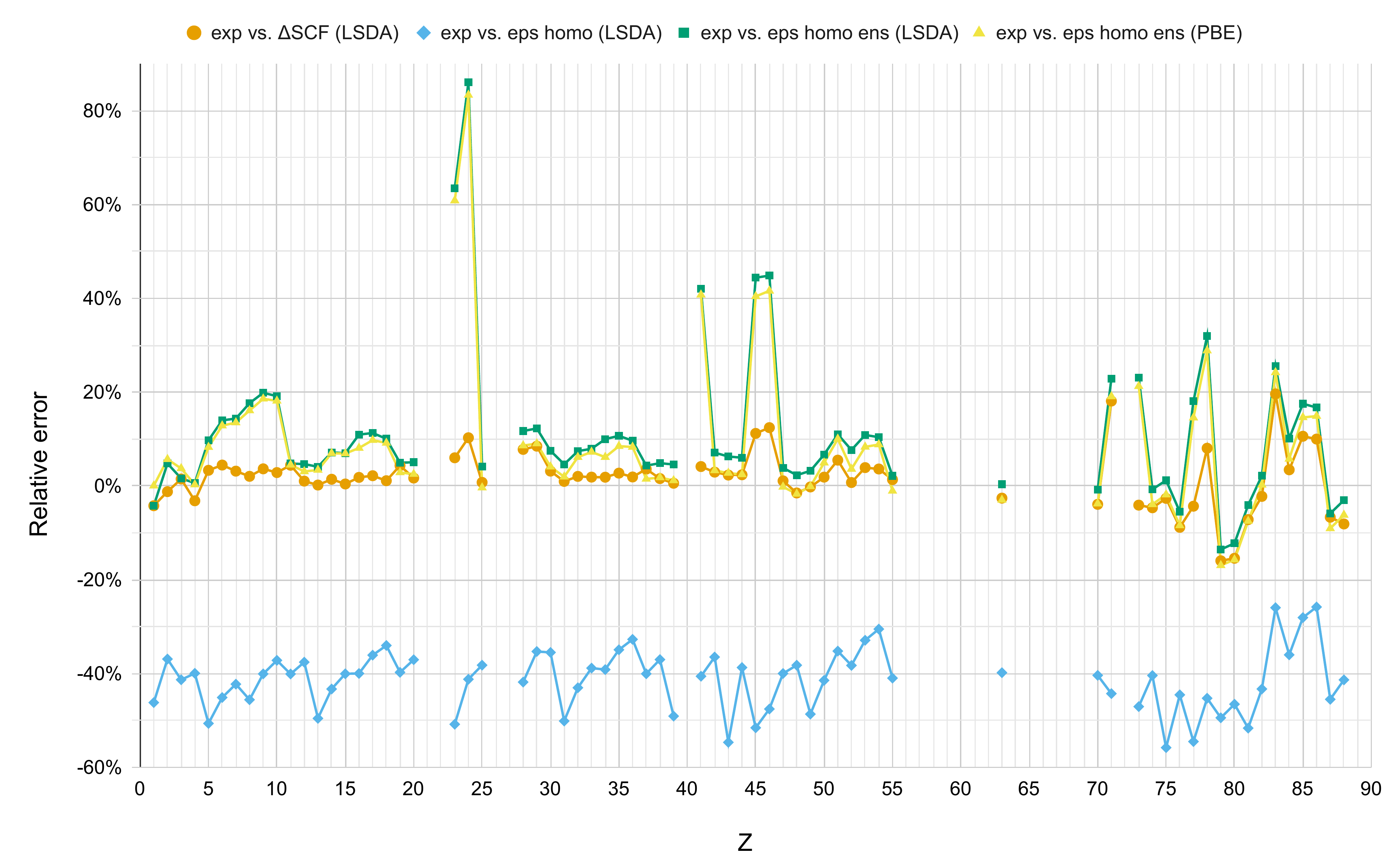}
\caption{The relative error $\delta_{\IP}$ in the ionization potential of neutral atoms with respect to experiment, as a function of the atomic number~$Z$, obtained from LSDA $\DSCF$ (orange circles), from LSDA homo energies before the ensemble correction (light blue rhombuses), and after the ensemble correction (green squares), and from PBE homo energies after the ensemble correction (yellow triangles).}
\label{fig:relative errors v0}
\end{figure*}

 Figure~\ref{fig:IP Vs.Z} presents the IP of neutral atoms with \mbox{$1 \leqs Z \leqs 88$}, obtained as the negative of the homo energy. Ensemble-corrected results are presented for LSDA ($-\tilde{\eps}_\ho^\LSDA$) and PBE ($-\tilde{\eps}_\ho^\PBE$). In addition, for comparison we present the LSDA homo values before the ensemble correction ($-\eps_\ho^\LSDA$), IPs calculated from LSDA total energy differences ($\IP_\DSCF^\LSDA$), and experimental IPs ($\IP_{\exp}$)~\cite{HandChemPhys95, Rothe2013, Finkelnburg1955, Chang2010}. Results for PBE homo energies before the ensemble correction ($-\eps_\ho^\PBE$) and PBE total energy differences ($\IP_\DSCF^\PBE$) are not presented for sake of clarity, because they are very close to the corresponding LSDA quantities. 
  Figure~\ref{fig:relative errors v0} presents the relative error in the IP with respect to the experimental values,  $\delta_{\IP} = (\IP_\textrm{calc}-\IP_\textrm{exp})/\IP_\textrm{exp}$, for each of the quantities of Fig.~\ref{fig:IP Vs.Z}. 
  Table~\ref{table:rel_error_IP} shows the average of the absolute relative errors, $|\delta_{\IP}|$, for each block of the Periodic Table.
  
  \begin{table}[h]
  \centering
    \begin{tabular}{|c|ccc|ccc|}
     \hline\hline
          &  \multicolumn{3}{c|}{LSDA} &  \multicolumn{3}{c|}{PBE} \\
    %\hline
    Block &  $\IP_\DSCF$ & $-\eps_\ho$ & $-\tilde{\eps}_\ho$ & $\IP_\DSCF$ & $-\eps_\ho$ & $-\tilde{\eps}_\ho$ \\
    \hline
%LSDA
%3.3%	40.3%	3.9% 
%3.8%	39.7%	11.3%
%6.0%	44.5%	19.7%

%PBE
%3.2%	41.2%	3.3%
%3.1%	40.9%	9.9%
%6.1%	47.6%	17.9%
    $s$   &  3.3\% & 40.3\% & 3.9\% & 3.2\% & 41.2\% & 3.3\% \\
    $p$   &  3.8\% & 39.7\% & 11.3\% & 3.1\% & 40.9\% & 9.9\% \\
    $d$   &  6\% & 44.5\% & 19.7\% & 6.1\% & 47.6\% & 17.9\% \\
    \hline
%4.5%	41.3%	12.3%
%4.3%	43.2%	10.9%
    overall & 4.5\% & 41.3\% & 12.3\% & 4.3\% &  43.2\% & 10.9\% \\
     \hline\hline
    \end{tabular}
    \caption{The average absolute relative errors of the ionization potential obtained within the LSDA and the PBE-GGA with respect to experiment, via three methods: from total energy differences ($\IP_\DSCF$), and from the homo before ($-\eps_\ho$) and after ($-\tilde{\eps}_\ho$) the ensemble correction}\label{table:rel_error_IP}
\end{table}

 From Figs.~\ref{fig:IP Vs.Z} and~\ref{fig:relative errors v0} and Table~\ref{table:rel_error_IP} we infer that the ensemble generalization, i.e.\ the transformation from $\eps^X_\ho$ to $\tilde{\eps}^X_\ho$ ($ X \in \{ \LSDA, \PBE \} $) brings a significant improvement to the prediction of the IP, across the Periodic Table. The overall average error reduces from 41.3\% to 12.3\% for the LSDA and from 43.2\% to 10.9\% for PBE. Thus, improved accuracy in the calculation of the IP from the homo energy can be obtained even if the underlying xc approximation is as simple as the LSDA. %These results are in agreement with the conclusions of Ref.~\cite{KraislerSchmidt15}. 
 
 The consistent underestimation of the IP values changes to a smaller and consistent overestimation (see Fig.~\ref{fig:relative errors v0}). This result is in agreement with the finding~\cite{KraislerKronik13, KraislerKronik15} that upon ensemble generalization the convex $E(N)$ curve usually becomes a concave one. 
 
 Furthermore, the homo values for the PBE functional are very close to those of LSDA, both before and after the ensemble treatment. When high errors are observed in $-\tilde{\eps}_\ho^\LSDA$, the PBE functional does not provide a remedy. These results are in agreement with calculations of Ref.~\cite{KraislerSchmidt15}, which were performed for a smaller set of systems, but with more xc functionals. 
 
 Interestingly, whereas the discrepancy of $-\eps^X_\ho$ (i.e., before the generalization) versus the experiment is very similar for all the blocks of the Periodic Table (circa 40\%; see Table~\ref{table:rel_error_IP}), the discrepancy of $-\tilde{\eps}^X_\ho$ (after the generalization) crucially depends on the position of the atom in the Table. Whereas for the $s$-block the average absolute relative error in $-\tilde{\eps}^X_\ho$ is 3.9\% for the LSDA and 3.3\% for the PBE, for the $p$-block the error is 11.3\% and 9.9\%, respectively, and for the $d$-block it is significantly higher: 19.7\% and 17.9\%, respectively. %\orange{Moreover, the relative error of $-\tilde{\eps}^X_\ho$ vs $\DSCF$ results, we find that $\delta_{\IP}$ for the $s$-block is 2.2\% for the LSDA and 2.3\% for the PBE, for the $p$-block the error is 7.3\% and 7.3\%, respectively, and for the $d$-block it is significantly higher: 14.6\% and 15.4\%, respectively.}
 This situation calls for an explanation. 
 Therefore, in Sec.~\ref{sec:Results:IP_d} we focus on the $d$-block to perform a more detailed analysis of the IPs.


\subsection{Ionization potentials of $d$-block atoms} \label{sec:Results:IP_d}
 
As mentioned in the previous section, the average absolute deviation of $-\tilde{\eps}_\ho$ from experiment is significantly higher for $d$-block atoms. A closer look at the data reveals that within the $d$-block there are atoms for which $\delta_{\IP}$ is rather small, e.g., Mn, Zn, Y, Mo, Tc, Ru, Ag and Cd ($Z=25$, 30, 39, 42, 43, 44, 47, and 48, respectively)~-- below 10\%. In parallel, for other atoms $\delta_{\IP}$ is much larger. It is largest for Cr ($Z=24$) with the LSDA, +86\%, but also V, Nb, Rh, Pd and Pt ($Z=23, 41, 45, 46$ and $78$) yield errors above 30\%. In these systems, there is also a prominent difference between $-\tilde{\eps}_\ho$ and the $\IP$ calculated from total energy differences, as can be seen in Fig.~\ref{fig:IP Vs.Z}.
The existence of this group made us wonder what is common to the systems where our approach fails. Apparently, in all these systems the homo is a $d$-orbital, whereas for systems with low relative errors the homo is an $s$-orbital. 

It is well known that in the $d$-block the $d$ and $s$ KS orbitals are very close in energy. Therefore, for some atoms in this block the homo is a $d$-state and for others -- an $s$-state. For example, in the Nb atom the difference between the 4$d^\up$ and the 5$s^\up$ energy levels is 0.0132 Hartree in the LSDA and 0.0097 Hartree in the PBE-GGA. For such small deviations, it is reasonable to expect that different xc approximations will not all yield 4$d^\up$ as the homo of Nb. In fact, xc functionals that aim to remove, or at least reduce the self-interaction error, are expected to lower the $d$-states and make 5$s^\up$ the homo~\cite{PZ81}. 

\begin{figure}
  \centering
  \includegraphics[width=0.49\textwidth]{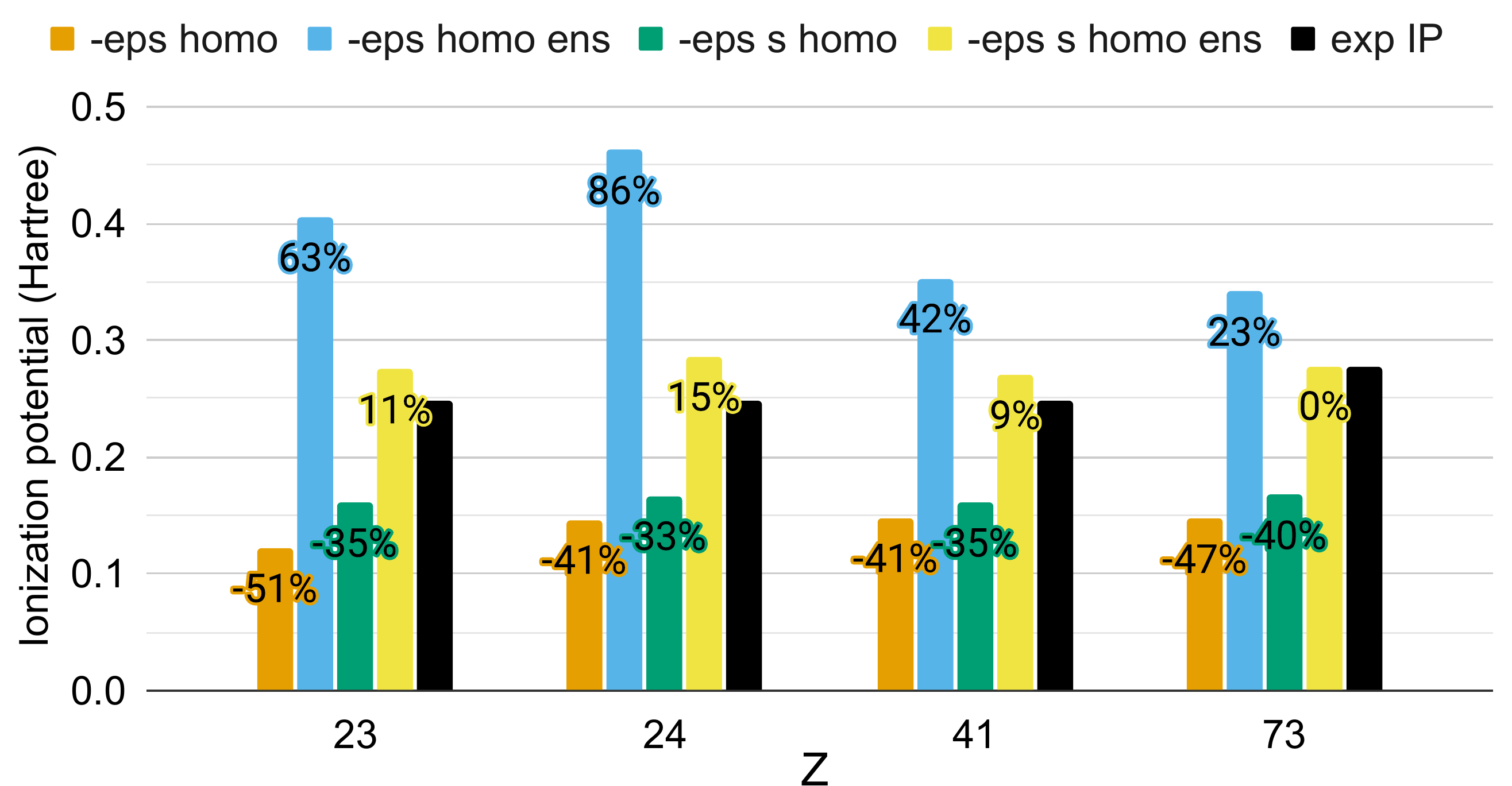}
%\caption{The relative errors in the ionization potential of V, Cr, Nb and Ta ($Z=23, 24, 41$ and $73$, respectively), using the original LSDA homo, before (orange) and after (light blue) the ensemble generalization correction, and the LSDA $s$-orbital energy level before (green) and after (yellow) the ensemble generalization correction, with respect to the experiment.}
\caption{
The ionization potentials of V, Cr, Nb and Ta ($Z=23, 24, 41$ and $73$, respectively), calculated using the original LSDA homo, before (orange) and after (light blue) the ensemble generalization correction, and the LSDA $s$-orbital energy level before (green) and after (yellow) the ensemble generalization correction. Experimental values (black) are shown for comparison. Each column is labeled by its relative error with respect to experiment.}
\label{fig:d block analysis}
\end{figure}

For this reason, we perform a heuristic analysis for four $d$-block atoms where the homo is a $d$-state: V, Cr, Nb and Ta ($Z=23, 24, 41, 73$). Since the $d$- and $s$-eigenvalues in these atoms are very close and the $s$-level is occupied, we take the liberty to declare the $s$-level to be the homo. We then recalculate the potential shift $v_0$ (Eq.~(\ref{eq:spatially uniform term})) accordingly, obtain new $\tilde{\eps}_\ho$ and compare to the experimental IP. Figure~\ref{fig:d block analysis} summarizes these results for the LSDA. Whereas before the ensemble treatment both the $d$- and the $s$-eigenvalues strongly underestimate the IP ($-30$\% to $-50$\%), after the ensemble treatment the $d$-levels yield significant overestimation that reaches 86\% for Cr. The $s$-eigenvalues perform much better, with an average overestimation of only 8.7\%. For Ta the deviation is essentially 0\%. Performing the same calculations with the PBE yields even better correspondence, with an average overestimation of only 5.5\%. 

\begin{figure}
  \centering
  \includegraphics[width=0.49\textwidth]{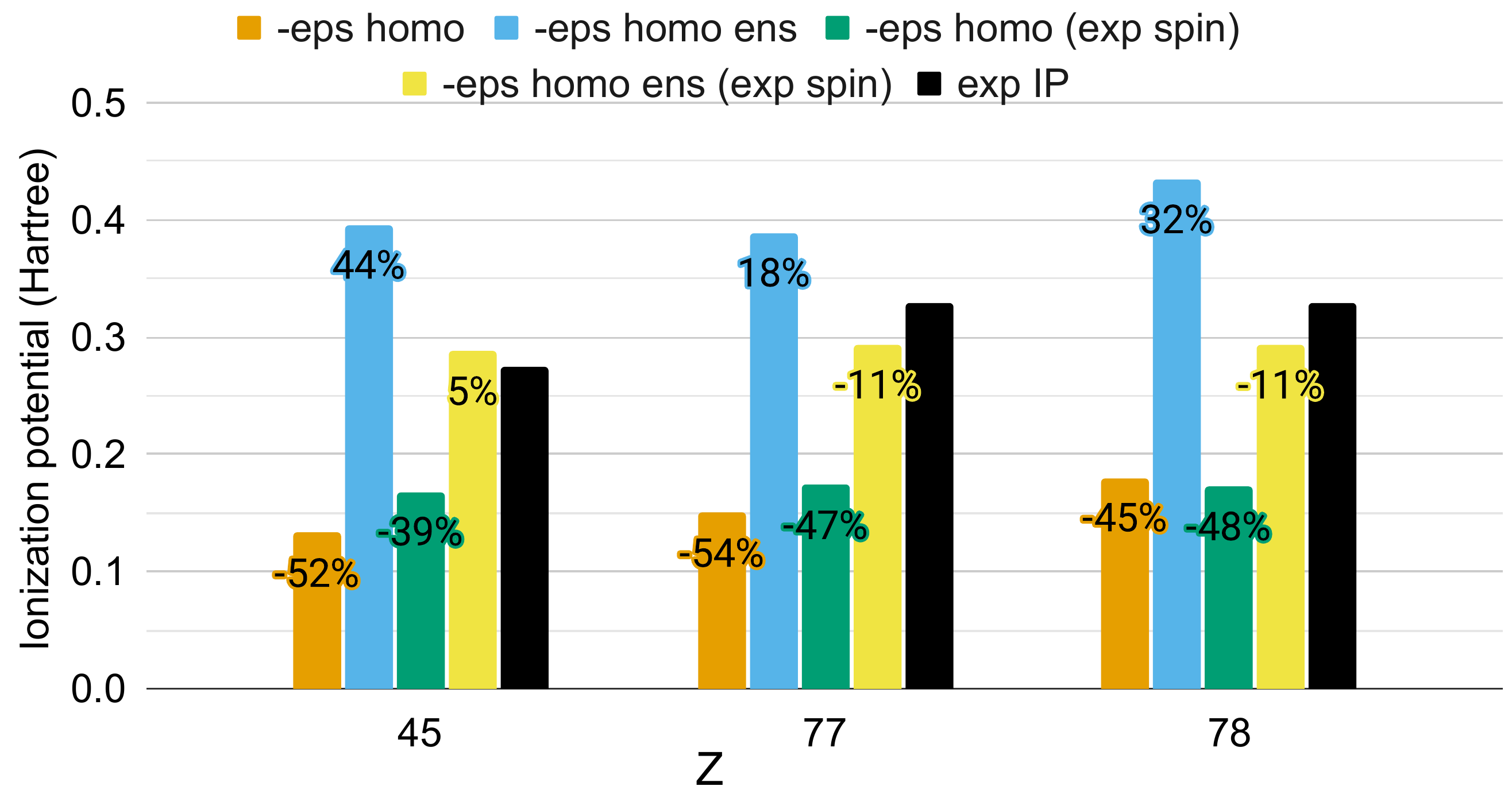}
\caption{The ionization potentials of Rh, Ir and Pt ($Z=45, 77$ and $78$, respectively), calculated using the original LSDA homo, before (orange) and after (light blue) the ensemble generalization correction, and the LSDA homo from the calculation done with the experimental value of the spin, before (green) and after (yellow) the ensemble generalization correction. Experimental values (black) are shown for comparison. Each column is labeled by its relative error with respect to experiment.}
\label{fig:d block analysis 2}
\end{figure}

For Rh, Pd, Ir and Pt ($Z=45,46,77$ and $78$), although the error between the IP and $-\tilde{\eps}_\ho$ is considerable, we could not perform this heuristic analysis, because the $s$-level is vacant. 
However, since for Rh, Ir and Pt ($Z=45,77$ and $78$) the total spin obtained with the LSDA and PBE (by minimizing the total energy $E(S)$) differs from the experimental value, we also examine the case where the spin is chosen to match the experiment. When we do this, the homo becomes an $s$-level, and $-\tilde{\eps}_\ho$ more closely approximates the IP. These results are summarized in Fig.~\ref{fig:d block analysis 2} for the LSDA.

Finally, for Pd ($Z=46$), where the spin obtained with the LSDA and PBE does match the experimental value, we artificially occupied the $5s$ level, at the expense of $4d$, and reran the calculation with the occupations frozen; 5$s$ then became the new homo level.
The ground state KS electronic configuration obtained for Pd is (i)~[Kr]4$d^5_5$, and we consider the occupations (ii)~[Kr]4$d^4_4$ 5$s^1_1$ and (iii)~[Kr]4$d^4_5$ 5$s^1_0$, which have higher total energies than (i) (with the LSDA, by 0.07 Hartree and 0.23 Hartree, respectively). %, but for the exact functional it may be that either (ii) or (iii) is the ground-state occupation. 
The values of $-\tilde{\eps}_\ho$ obtained with these occupations deviate from the experimental IP by (i): 44.8\%, (ii): $-7.9$\% and (iii): 6.3\%, respectively. 
%In sight of this improvement, it is plausible that the occupation numbers obtained with the LSDA and PBE are incorrect for Pd, and that for the exact xc functional, the $5s$ level in Pd really is occupied and is the homo.

These three examples, where, under different circumstances, the homo becomes an $s$ level, instead of a $d$ level, emphasize the sensitivity of our correction to the character of the homo: \emph{if an $s$-level, the prediction of the IP is much more accurate}.

In addition, in the $d$-block we found 7 atoms for which the homo level changed as a result of the ensemble generalization, from one spin channel to the other. As we mentioned in Sec.~\ref{sec:Theoretical Background}, $\tilde{\eps}_\ho$ is chosen to be the highest between $\tilde{\eps}_\ho^\up$ and $\tilde{\eps}_\ho^\dn$. This means that the spin channel to which the global homo belongs can change upon the ensemble generalization. This is exactly what happens in Ni, Y, Tc, Ru, Lu, Re and Os ($Z=28$, 39, 43, 44, 71, 75 and 76), in both the LSDA and the PBE. In all these cases the homo changed from a $d$-level to an $s$-level after the ensemble correction. Choosing the global homo after the ensemble generalization anew, and not remaining with the previous choice for the homo, brings a dramatic improvement in these atoms. For example, for Ni within the LSDA, an error of 107.5\% in $\tilde{\eps}_\ho$ is obtained when not updating the choice of the global homo; it is reduced to 11.7\% when doing so. This result further strengthens our observation above that when the homo is a $d$-level, a higher error is expected.

Finally, in the context of $d$-block atoms, we notice the relatively high underestimation of the IP for Au and Hg ($Z=79$, 80) after the ensemble treatment: 12-17\%. But in these atoms also the IP obtained from total energy differences largely underestimates the experiment, due to relativistic effects that are absent from our calculations. Thus, the ensemble correction is not expected to improve results in these atoms: it corrects $\eps_\ho$ to correspond better to $\IP_\DSCF$, but if the $\IP_\DSCF$ is in error, the ensemble treatment cannot do better.

\subsection{Fundamental gaps of first ions with $1 \leqs Z \leqs 88$ } \label{sec:Results:Eg_ions}

In this Section we evaluate the fundamental gaps of the first ions using the LSDA and PBE-GGA. We chose to first focus on the ions because most (semi-)local density functionals do not bind atomic anions due to large self-interaction errors~\cite{Primer}. Therefore, the EA is not available computationally, and comparison to $\DSCF$ results is not possible. In contrast, the fundamental gap of all first ions is easily calculated as $\IP_2 - \IP_1$, the difference of the second and the first ionization potentials. Furthermore, the corresponding experimental data %on gaps of ions 
is abundant and highly accurate.

\begin{figure}
  \centering
  \includegraphics[width=0.47\textwidth]{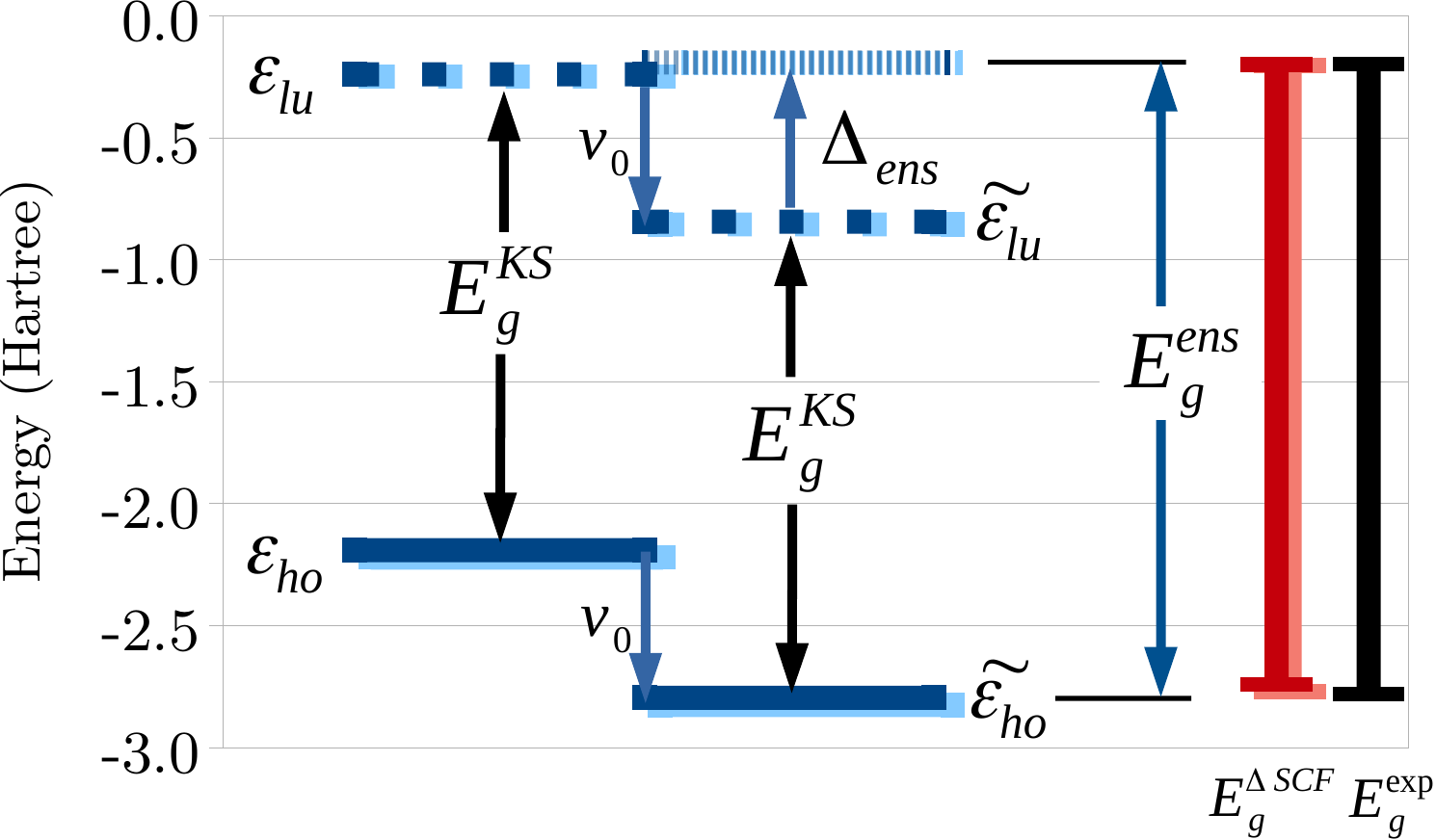}
\caption{Highest occupied and lowest unoccupied energy levels of Li$^+$ before ($\eps_\ho$ and $\eps_\lu$) and after the ensemble generalization ($\tilde{\eps}_\ho$ and $\tilde{\eps}_\lu$). The potential shift $v_0$ is depicted, as well as the derivative discontinuity $\Delta_\ens$, which adds up to the KS gap, $E_g^\KS$, to form the fundamental gap, $E_g^\ens$. LSDA (dark blue) and PBE (light blue) values are almost indistinguishable. Values for the fundamental gap obtained from total energy differences, $E_g^\DSCF$ (LSDA - dark red, PBE - light red) and from experiment $E_g^{\exp}$ (black) are plotted for comparison.}
\label{fig:Eg_Li+}
\end{figure}

\begin{figure}
  \centering
  \includegraphics[width=0.47\textwidth]{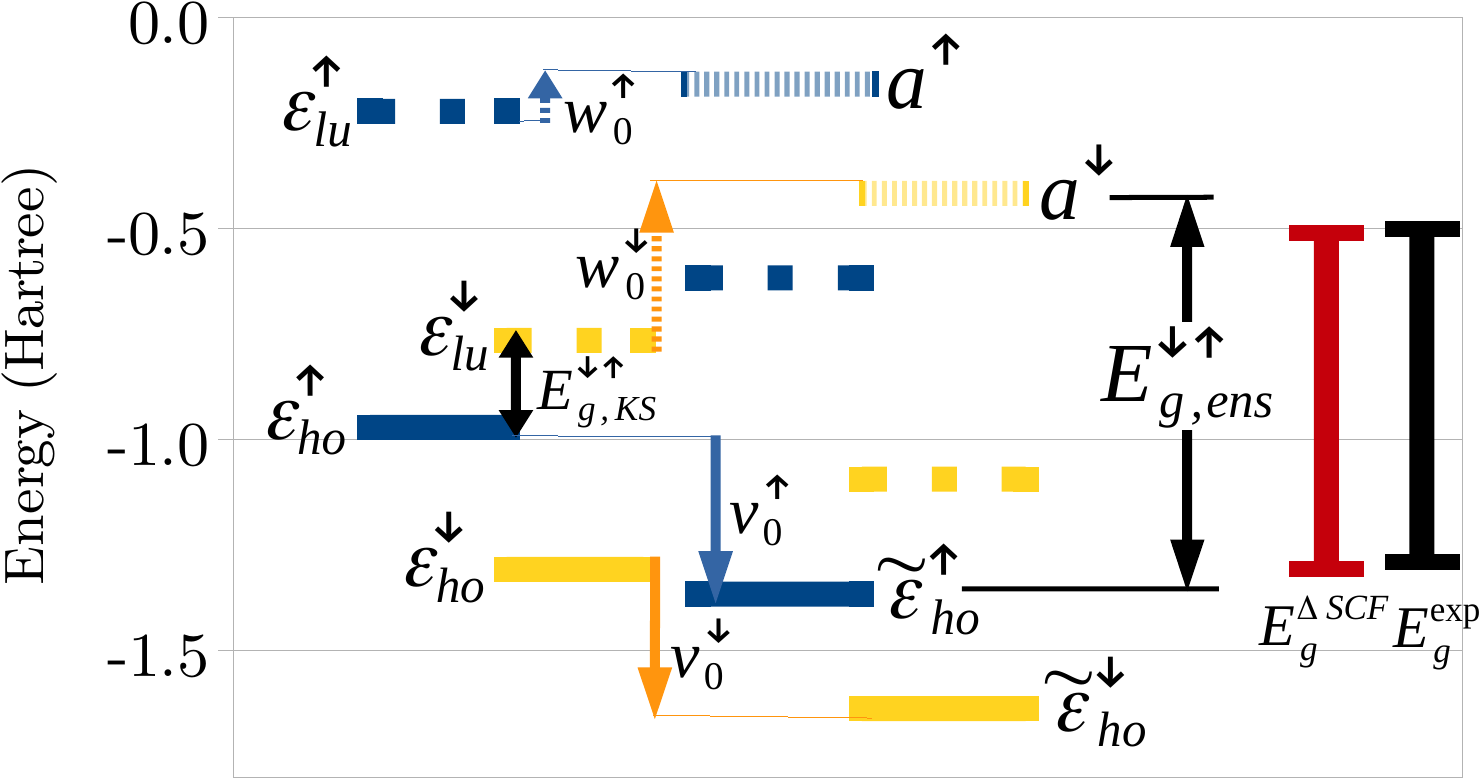}
\caption{Highest occupied and lowest unoccupied energy levels of O$^+$ ($\eps_\ho^\up$ and $\eps_\lu^\up$ -- solid and dotted blue and $\eps_\ho^\dn$ and $\eps_\lu^\dn$ -- solid and dotted yellow)  within the LSDA before and after the ensemble generalization. The potential shifts $v_0^\s$ are depicted by solid arrows and the shifts $w_0^\s$ by dashed arrows. The resultant energies $a^\up$ and $a^\dn$ are drawn in dashed blue and yellow, respectively. The smallest KS gap, $E_{g,\KS}^{\dn \up}$ and the smallest fundamental gap, $E_{g,\ens}^{\dn \up}$ are indicated. Values for the fundamental gap obtained from total energy differences, $E_g^\DSCF$ (red), and from experiment, $E_g^{\exp}$ (black), are plotted for comparison.}
\label{fig:Eg_O+}
\end{figure}

We first illustrate our approach on two systems -- Li$^+$ and O$^+$, in Figures~\ref{fig:Eg_Li+} and~\ref{fig:Eg_O+}, respectively. For Li$^+$ there are two electrons, one in each spin channel. The $\up$- and $\dn$-levels in this system are identical, and therefore are not plotted separately. Prior to the ensemble treatment, with the LSDA, $\eps_\ho = -2.190$~Hartree and $\eps_\lu = -0.240$~Hartree, which yields a KS gap of $E_g^\KS= 1.950$~Hartree. The experimental gap for Li$^+$ is $E_g^{\exp} =2.582$~Hartree, %i.e.\ the LSDA underestimates the experimental figure by 24\%. 
and the gap obtained from the LSDA total energy differences is $E_g^\DSCF= %\IP(\textrm{Li}^+) - \EA(\textrm{Li}^+) =\IP_2(\textrm{Li})-\IP_1(\textrm{Li})=
2.539$~Hartree. We therefore see that $E_g^\DSCF$ is able to closely approximate $E_g^{\exp}$, but $E_g^\KS$ underestimates it by $24\%$. 
PBE results are very close for this case, almost indistinguishable in Fig.~\ref{fig:Eg_Li+}, and therefore are not discussed at length.

Due to the ensemble generalization, all energy levels are shifted by $v_0=-0.605$~Hartree. In addition, a derivative discontinuity $\Delta_\ens = 0.652$~Hartree adds up to the KS gap to form the fundamental gap $E_g^\ens = 2.602$~Hartree. It favourably compares to $E_g^\DSCF$, with 2.5\% overestimation, and to the experimental gap $E_g^{\exp}$, with only 0.8\% overestimation. Thus, the ensemble generalization brings a significant improvement in predicting the gap for Li$^+$.

There are two equivalent ways to look at the results depicted in Fig.~\ref{fig:Eg_Li+}: (i) all energy levels move by the same potential shift, $v_0$; this does not change $E_g^\KS$. The discontinuity $\Delta_\ens$ adds up to $E_g^\KS$ to form $E_g^\ens$.  (ii) $\eps_\ho$ is shifted by $v_0$ and becomes $\tilde{\eps}_\ho = \eps_\ho + v_0$; it can then be compared to $-\IP$
 (in the case of an ion, $-\IP_2$). $\eps_\lu$ is first also shifted by $v_0$ but then again is shifted by $\Delta_\ens$ to yield the quantity $a = \tilde{\eps}_\lu + \Delta_\ens$; it can then be compared to $-\EA$ (in the case of an ion, to $-\IP_1$). This quantity can also be expressed as $a = \eps_\lu + w_0$, where $w_0 = \Delta_\ens + v_0 = E_\Hxc[n+{|\pphi_{\lu}|}^{2}] - E_\Hxc[n] - \int  |\pphi_{\lu}(\rr)|^{2} v_\Hxc[n](\rr) \, d^{3}r$ (cf.\ Eqs.~(\ref{eq:delta ens}) and~(\ref{eq:spatially uniform term})).
  The difference $a-\tilde{\eps}_\ho$ equals the fundamental gap, $E_g^\ens$. This second perspective proves particularly useful in spin-dependent cases, like O$^+$, which is discussed below.

In the case of O$^+$, the spin channels are not equivalent and therefore the diagram of energy levels is more complex (Fig.~(\ref{fig:Eg_O+})). 
Within the LSDA, in the $\up$-channel, $\eps_\ho^\up = -0.971$~Hartree and $\eps_\lu^\up = -0.222$~Hartree. In the $\dn$-channel, $\eps_\ho^\dn = -1.308$~Hartree and $\eps_\lu^\dn = -0.765$~Hartree. %Therefore, the global homo is $\eps_\ho^\up$ and the global lumo is $\eps_\lu^\dn$. 
From these four frontier orbitals, one can in principle get four KS gaps: $E_{g,\KS}^{\tau \s} := \eps_\lu^\tau - \eps_\ho^\s$, with $\s, \tau \in \{ \up, \dn \}$ (notice that the order of the indices $\s$ and $\tau$ matters!). To compare to experiment, we choose the smallest gap of these four:
\begin{align}
    E_g^\KS = \min_{\tau \s} E_{g,\KS}^{\tau \s} = \min_\tau ( \eps_\lu^\tau) - \max_\s (\eps_\ho^\s)
\end{align} 
In other words, the minimal KS gap is obtained as the difference between the global lumo and the global homo energies. In the case of O$^+$, $E_g^\KS = E_{g,\KS}^{\dn \up} = 0.207$~Hartree. This result underestimates the experimental gap by 74\%  ($E_g^{\exp} =0.79$~Hartree) as well as the gap obtained from LSDA total energy differences ($E_g^\DSCF=0.796$~Hartree). 
Due to the ensemble generalization, the levels in each spin channel are shifted by $v_0^\up = -0.395$~Hartree and $v_0^\dn = -0.329$~Hartree, respectively. This yields the ensemble-corrected homo levels $\tilde{\eps}_\ho^\up = -1.366$~Hartree and $\tilde{\eps}_\ho^\dn = -1.637$~Hartree. Next, in principle, one needs to calculate %\red{two DD's for each level: $\Delta_\ens^{\up \up}$ and $\Delta_\ens^{\dn \up}$ for $\tilde{\eps}_\ho^\up$ and $\Delta_\ens^{\up \dn}$ and $\Delta_\ens^{\dn \dn}$ for $\tilde{\eps}_\ho^\dw$, obtain} 
the four fundamental gaps, $E_{g,\ens}^{\tau \s}=E_{g,\KS}^{\tau \s}+\Delta_\ens^{\tau \s}$ and find the smallest one. Alternatively, and completely equivalently, one can calculate the quantities 
\begin{align} \label{eq:w_0_tau}
w_0^\tau = v_0^\s + \Delta_\ens^{\tau \s} &= E_\Hxc[n+{|\pphi^\tau_{\lu}|}^{2}] - E_\Hxc[n] \nonumber \\
& - \int  |\pphi^\tau_{\lu}(\rr)|^{2} v_\Hxc^\tau (\rr) \, d^{3}r,    
\end{align}
which notably depend only on $\tau$, but not on $\s$. Then, one obtains $a^\tau = \eps_\lu^\tau + w_0^\tau$ (see Fig.~\ref{fig:Eg_O+}) and gets the gap by choosing the highest $\tilde{\eps}_\ho^\s$ and the lowest $a^\tau$:  $E_g^\ens = \min_\tau (a^\tau) - \max_\s (\tilde{\eps}_\ho^\s)$. For O$^+$, $w_0^\up = 0.064$ Hartree and $w_0^\dn = 0.348$~Hartree. Then, the lowest $a$ is $a^\dn = -0.417$~Hartree and the ensemble-corrected fundamental gap is $E_g^\ens = E_{g,\ens}^{\dn \up} = 0.949$~Hartree. The deviation from experiment is therefore reduced from -74\% to +20\%.
Importantly, the spin indices of the minimizing $E_g^\ens$ and $E_g^\KS$ do not have to be the same, and the optimization vs.\ $\s$ and $\tau$ must be redone after the ensemble treatment. Cases when this is important are mentioned below.

\begin{figure*}
  \centering
  \includegraphics[width=1.0\textwidth]{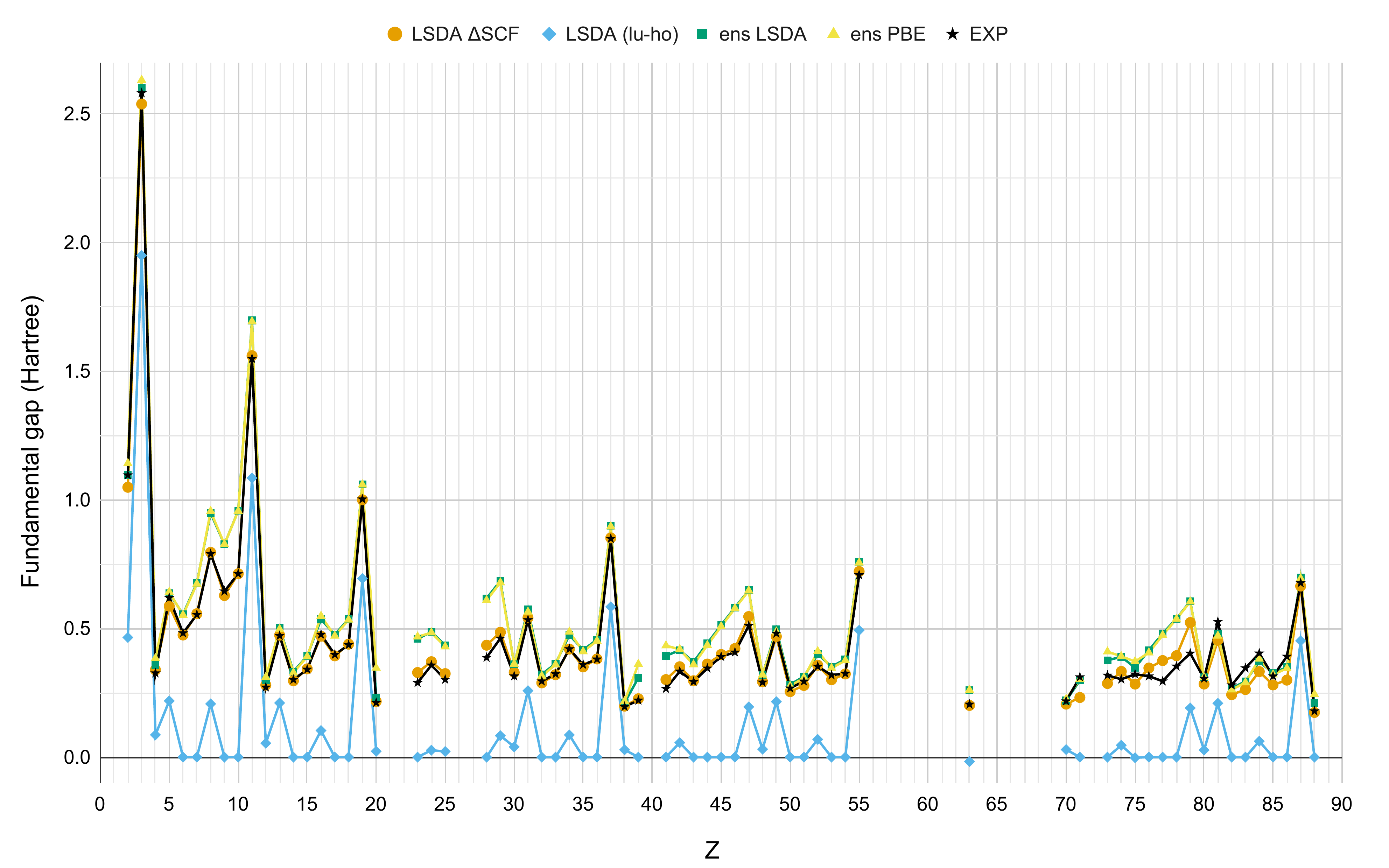}
\caption{The fundamental gap of first ions, as a function of the atomic number $Z$, for the LSDA before (light blue rhombuses), and after (green squares) the ensemble correction, and
for the PBE after the ensemble correction only (yellow triangles). Experimental values (black stars) and LSDA  $\DSCF$ results (orange circles) are shown for comparison.}
\label{fig:fg}
\end{figure*} 

\begin{figure*}
  \centering
  \includegraphics[width=1.0\textwidth]{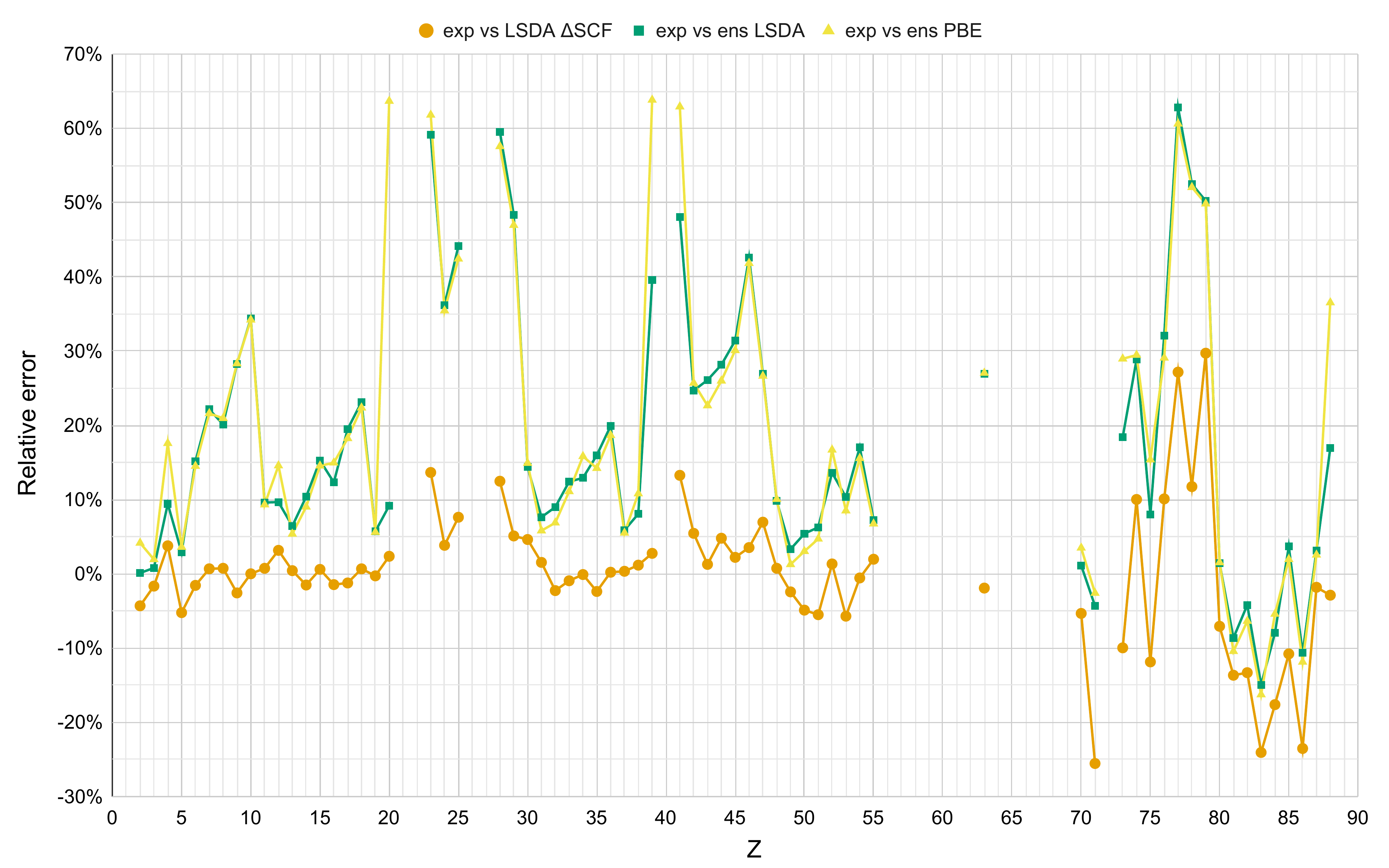}
\caption{The relative errors in the fundamental gap of first ions, with respect to experiment, as a function of the atomic number~$Z$, obtained from LSDA $\DSCF$ (orange circles), LSDA ensemble-corrected gaps $E_g^\ens$ (green squares), and PBE ensemble-corrected gaps (yellow triangles).}
\label{fig:relative erros Eg}
\end{figure*}

After a detailed analysis of two particular examples, we present results for all the first ions we calculated, in Figs.~\ref{fig:fg} and~\ref{fig:relative erros Eg}.
Overall, the improvement in the prediction of the fundamental gap in ions is dramatic: $E_g^\KS$ systematically underestimates the experiment; the average absolute error with the LSDA is 85\%, and with the PBE -- 84\%. Introducing the ensemble DD, the average error with the LSDA becomes 19\%, and with the PBE -- 21\%. With the exception of a few cases, the ensemble-generalized functionals overestimate the gap.

The ensemble correction is of particular importance in those systems where the KS gap is 0, i.e.\ when $\eps_\ho = \eps_\lu$. This situation is very common: out of 68 ions considered, 34 have zero KS gap %\blue{and 2 have a negative KS gap} 
with the LSDA. For these systems, the whole burden of reproducing the fundamental gap is born by the derivative discontinuity.
%\blue{For these systems, a prediction of the fundamental gap is enabled by the ensemble DD.}
Calculated now as $\Delta_\ens$ of Eq.~(\ref{eq:delta ens}), it was completely absent before. %Underestimation of 100\% becomes an overestimation of 24\% with LSDA and 25\% with PBE.
In these cases, where the KS gap underestimates the fundamental gap by 100\%, the average absolute error after the ensemble correction is 24\% with the LSDA and 25\% with the PBE.

Also in cases when $E_g^\KS$ is not zero, the contribution of $\Delta_\ens$ is always substantial, and cannot be neglected. With the LSDA, the ratio $|\Delta_\ens/E_g^\KS|$ is lowest for Li$^+$ with 0.334, and largest for Re$^+$ with 161 (due to a very low KS gap in this ion).

%\orange{*** The improvement in the prediction of the fundamental gap is dramatic: average; consistent overestimation. Particularly: systems with zero KS gap.}

%\orange{*** Analyzing by groups: both levels are s, both are p, ... d block in general, within d-block: both levels are s ...}
Next, we consider the orbitals used to calculate $E_g^\ens$: $\tilde{\pphi}_\ho \equiv \pphi_\ho^\s$ and $\tilde{\pphi}_\lu \equiv \pphi_\lu^\tau$, where $\s$ is the spin which maximizes $\tilde{\eps}_\ho^\s$ and $\tau$ minimizes $a^\tau$ (and not $\tilde{\eps}_\lu^\tau$). We stress that the spatial dependence of \emph{all} the orbitals, $\pphi_{i\s}(\rr)$, does not change due to the ensemble correction, at integer $N$. What does change is the eigenvalues, $\eps_{i\s}$, and hence their ordering. As a result, the characterization of a given orbital as the homo or lumo changes, as well. Note that $\tilde{\pphi}_\ho$ \emph{is} the homo obtained after the ensemble correction, but $\tilde{\pphi}_\lu$ is not necessarily the lumo before or after the ensemble correction -- it is the lumo of the spin channel $\tau$, which minimizes $a^\tau$. 
Analyzing the results by the character of $\tilde{\pphi}_\ho$ and $\tilde{\pphi}_\lu$, we find that in cases where either of them is a $d$-level, the average deviation of $E_g^\ens$ is much higher: 35\% in the LSDA and 39\% in the PBE, comparing to 12\% for all the other systems in both functionals.
This further strengthens our observation regarding the lower accuracy of the ensemble correction for $d$-levels.

Importantly, there are cases where $\pphi_\lu$ (before the ensemble correction) differs from $\tilde{\pphi}_\lu$, meaning the first spin index of the minimizing $E_g^\KS$ is different from that of $E_g^\ens$. This occurs in 19 systems with the LSDA ($Z=20$,23,25,28,39,41,43,44,45,46,71,73,76,77 and 88) and in 11 systems with the PBE ($Z=4$,23,25,28,44,45,46,63,71,76 and 77). In contrast, the global homo (and correspondingly, the second spin index of the minimizing gap) remains the same after the ensemble correction in \emph{all} of the first ions considered, as opposed to the situation with the neutrals (see Sec.~\ref{sec:Results:IP_d}).

Furthermore, certain atoms and ions are known to have a KS electronic configuration that is \emph{proper in the broad sense}~\cite{KraislerMakovKelson10}. Namely, whereas each of the $\up$- and $\dw$-subsystems is occupied \emph{properly}, according to the \emph{Aufbau} principle, combining the two KS systems together may lead to an electronic configuration where, say, an occupied $\up$-level is higher than a vacant $\dw$-level~\footnote{It is emphasized that broad-sense proper configurations obey all the required
restrictions related to a rigorous definition and differentiability of energy functionals~\cite{KraislerMakovArgamanKelson09,KraislerMakovKelson10}. Altering the system's spin $S$ by an integer does not change the status of these systems, and only consideration of non-trivially fractional values of $2S$ may result in a proper configuration; this option has been precluded in previous works as well as here, and the spin was always a half-integer.}. As a result, a \emph{negative} KS gap is obtained for these systems.  For systems considered here this is the case for Eu$^+$ ($-0.017$ Hartree) and Re$^+$ ($-0.002$ Hartree) in the LSDA and Tc$^+$ ($-0.005$ Hartree) and Eu$^+$ ($-0.008$ Hartree) in the PBE.  In absence of any DD, in these atoms the fundamental gap is estimated to be negative, which is in contradiction to any experimental data~\footnote{It contradicts the so-called convexity conjecture for Coulomb systems, namely that $\IP > \EA$, for \emph{any system}~\cite{DG,Lieb,Cohen12,PPLB82}}.  Fortunately, by inclusion of $\Delta_\ens$, the fundamental gap of \emph{all} these systems is predicted to be positive (0.2602, 0.3479, 0.3597 and 0.2605 Hartree, respectively). Therefore, the problem of the negative gap for broad-sense proper systems is solved here by the ensemble generalization.

In general, also for the first ions the differences between the LSDA and PBE are minute.
In some systems, however, larger differences appear (see Fig.~\ref{fig:relative erros Eg}), which can be related, in many cases, to the fact that the LSDA and PBE identify the homo and the lumo differently.
%either $\tilde{\pphi}_\ho$ or $\tilde{\pphi}_\lu$ differently, meaning they result in different spin indices for the minimizing $E_g^\ens$. 
For the homo, this happens in Re$^+$ ($Z=75$), both before and after the ensemble correction. For the lumo, before the ensemble correction, this happens in Be$^+$, Tc$^+$ and Re$^+$ ($Z=4,43$ and $75$). After the ensemble correction, $\tilde{\pphi}_\lu$ is identified differently in Ca$^+$, Y$^+$, Nb$^+$, Ta$^+$, Re$^+$ and Ra$^+$  ($Z=20$, 39,41,73,75,88) after it.

\subsection{Fundamental gap calculations of neutral atoms} \label{sec:Results:Eg_neutral}

\begin{figure*}
  \centering
  \includegraphics[width=1.0\textwidth]{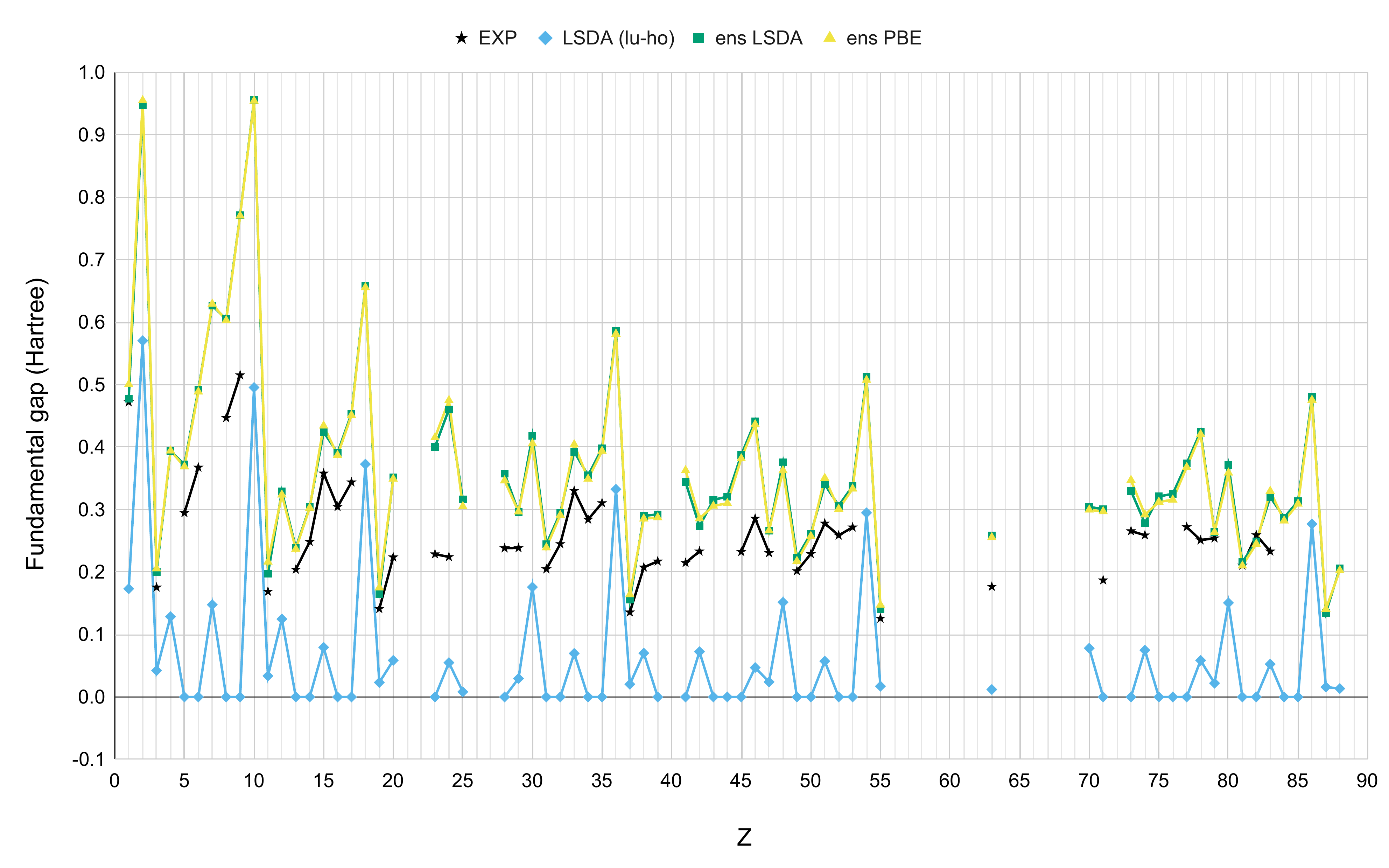}
\caption{The fundamental gap of neutral atoms obtained from the LSDA before (light blue rhombuses) and after (green squares) the ensemble correction, and from the PBE only after the ensemble correction (yellow triangles). Experimental values (black stars) are shown for comparison.}
\label{fig:neutrals_Eg}
\end{figure*}

Following our analysis of the fundamental gap for first ions, we now turn to analyze neutral atoms. 

The LSDA and PBE wrongly predict nearly all anions to be \emph{unbound}. Therefore, getting the fundamental gap from $\DSCF$ calculations becomes impossible. In contrast, $\Delta_\ens$ can still be calculated from Eq.~(\ref{eq:delta ens}) relying on quantities of the neutral atom, avoiding the calculation of the anion.

Results for the fundamental gap for neutral systems are presented in Fig.~\ref{fig:neutrals_Eg}.  Comparing to experiment, for the 47 atoms where experimental data exists~\cite{HandChemPhys95, Rothe2013, Finkelnburg1955, Chang2010}, we find that overall $E_g^\KS$ strongly underestimates the experimental gap, by 91\% with the LSDA and by 88\% with the PBE. Introducing $\Delta_\ens$, the average error becomes 31\% with both functionals. %with the LSDA becomes  19\%, and with the PBE -- 22\%.
With the exception of a few cases, the result obtained with the ensemble generalization overestimates the gap.

For 32 out of 69 atoms the KS gap was 0, and the gap results entirely from the ensemble correction. For neutrals, as for first ions,  $\Delta_\ens$ is never a negligible quantity: the ratio $|\Delta_\ens/E_g^\KS|$ ranges between 0.66 for He and 37.6 for Mn. As in first ions, when $d$-orbitals are involved, either as $\tilde{\pphi}_\ho$ or $\tilde{\pphi}_\lu$, the average error of $E_{g}^\ens$ is significantly larger: 56\% vs.\ 21\% for the other systems in the LSDA, and 57\% vs.\ 22\% in the PBE.

As we have discussed in Sec~\ref{sec:Results:IP_d}, the global homo changes as a result of the ensemble generalization in Ni, Y, Tc, Ru, Lu, Re and Os ($Z=28$, 39, 43, 44, 71, 75 and 76) within the LSDA and PBE. $\tilde{\pphi}_\lu$ differs from $\pphi_\lu$ in 14 systems within the LSDA ($Z=1$, 7, 8, 23, 25, 28, 41, 43, 44, 45, 63, 73, 75, 76, 77) and in 16 systems with the PBE (all those of LSDA and $Z=3$ and 15).

As with first ions, the differences in results with PBE vs.\ LSDA are small. There are some systems where $\tilde{\pphi}_\ho$ and $\tilde{\pphi}_\lu$ differ between these approximations: for $\tilde{\pphi}_\ho$ this is the case in Sr ($Z=38$) only, and for $\tilde{\pphi}_\lu$ in P, Sr, Xe and Ra ($Z=15,38,54,88$). 
%\red{***standard lumo vs ens lumo (minimizes $\tilde{\eps}_\lu$) vs $\tilde{\pphi}_\lu$?}

%\red{***PBE vs LSDA}

The fact that within a given xc approximation the anion is not bound, but the fundamental gap can still be obtained via the ensemble correction may look as an inconsistency in the theory: Predicting the system to be unbound means that the last electron is not attached to the nucleus and travels to infinity, which means that the energy of the anion equals that of the neutral atom, i.e., $\EA=0$ and $E_g$ should equal the IP. However, $E_g^\KS + \Delta_\ens$ does not necessarily equal the IP in this case. We address this point below. First, note that the results in this section can be interpreted according to the explanation above: from $E_g^\KS + \Delta_\ens$ and $\tilde{\eps}_\ho$ one can retrieve the quantity $a$ that should be compared to the EA, i.e., to 0. Analyzing the results this way yields that the average (signed) deviation of $a$ from 0 is 0.004 Hartree for the LSDA, which is a rather accurate result (highest deviation: 0.11 Hartree for Ca, lowest: -0.09 Hartree for At).  Second, note that with other approximations that do bind anions, e.g.\ hybrids, usage of Eq.~(\ref{eq:delta ens}) to obtain the fundamental gap would be totally legitimate. One may then claim for the LSDA and PBE that while the $\DSCF$ fails entirely for anions, the ensemble approach has the ability to predict the EA and hence the gap, which is one of its advantages.

\section{Discussion} \label{sec:Summary}

%\red{In this work, we applied the ensemble generalization procedure~\cite{KraislerKronik13} to atomic systems -- atoms and first ions across the Periodic Table -- aiming to obtain the IP and the fundamental gap departing from the KS spectrum. The underlying xc approximations which we generalized were the LSDA and PBE; however, our generalization is applicable to \emph{any} xc approximation. The additional numerical effort required by our approach is negligible. By applying our approach on a large scale, for systems with $s$-, $p$- and $d$-electrons, we were able to assess the accuracy of the method and identify important trends.}

In Sec.~\ref{sec:Results} we showed that the ensemble correction to the KS energy levels (Eqs.~(\ref{eq:Hxc potential generalization}), (\ref{eq:Delta_ens_tau_sigma}) and (\ref{eq:w_0_tau})) significantly improves the prediction of the IP and the fundamental gap for both neutral atoms and first ions, even if the underlying xc functionals are as simple as the LSDA or the PBE. The differences between PBE and LSDA results, after the ensemble generalization, were found to be minor. 

Analysis of the results clearly shows that the accuracy of our approach crucially depends on the position of the system in the Periodic Table, more precisely -- on the character of its frontier orbitals. In IP calculations, if the homo level is an $s$-level, the accuracy is found to be much higher than if it is a $d$-level (Sec.~\ref{sec:Results:IP}). This statement has been supported by the analysis performed in Sec.~\ref{sec:Results:IP_d}, where we heuristically re-identified the homo for selected transition metal systems. A similar trend was identified in our results for fundamental gaps (Secs.~\ref{sec:Results:Eg_ions} and~\ref{sec:Results:Eg_neutral}).

In our calculations of the fundamental gap for first ions we found a significant improvement achieved by the ensemble correction (Sec.~\ref{sec:Results:Eg_ions}). 
Furthermore, for neutral atoms the fundamental gap could be predicted with acceptable accuracy (Sec.~\ref{sec:Results:Eg_neutral}), even when the underlying xc approximations were as simple as the LSDA or PBE, while the usual route of total-energy differences was unavailable: the LSDA and PBE do not bind most anions.

Importantly, in our calculations we found that the derivative discontinuity $\Delta_\ens$ is always significant compared to the KS gap, and cannot be omitted. $\Delta_\ens$ is particularly significant in systems where the KS gap is zero, as it provides the whole fundamental gap. These systems can be viewed as the analogues of Mott insulators in atomic systems.

As to the failure for $d$-levels, we bring three possible reasons. First, the imperfections of the underlying xc functional, particularly the spurious self-interaction, may have caused the $d$-level to be too high, thus appearing as the homo instead of the competing $s$-level. 
%\orange{As we showed in Sec.~\ref{sec:Results:IP_d}, change in the levels' order significantly improves the prediction of the IP.}
As we showed in Sec.~\ref{sec:Results:IP_d}, the character of the homo substantially impacts the prediction of the IP.

Second, which partly explains also the moderate success for $p$-levels, the spherical approximation employed in the \texttt{ORCHID} code, serves as an error source in systems with open $p$- and $d$-subshells: a full 3D treatment of these systems, with the possibility to occupy the various $m$-levels differently and choose the KS ground state to be a pure state, may lead to a more accurate result within the xc approximations we used. This could also contribute to the levels' reordering and affect the ensemble shifts. 

The third reason relates to the nature of the KS ground states in $p$- and $d$-systems. Formulation of an additional generalization of the Hxc functional, which takes into account the fact that there exist degenerate levels in the KS potential, could be beneficial for further improvement of the results in $p$- and $d$-systems with open shells. All the aforementioned sources of error are desirable directions for future research. %This would be particularly relevant to systems with open $p$- and $d$-shells.

%\orange{Lastly, the third and deeper reason, closely related to the second one, is the fact that in $d$- and $p$-systems one of the assumptions of the ensemble generalization of Ref.~\cite{KraislerKronik13} may have not been satisfied: it has been originally assumed that the ground states of both the $(N_0-1)$-system (e.g.\ the cation) and the $N_0$-system (the neutral) are described by \emph{pure states}. Systems with $p$ and $d$ electrons, when treated three-dimensionally, experience a degenerate ground state. This was not taken into account in the original ensemble generalization. In the spherical approximation used here, the aforementioned degeneracy is not apparent, as all $d$- or $p$-states (of the same $n$) are treated as one state (with possible multiple occupancy).}
%\blue{Lastly, a third reason is that the ensemble generalization of Ref.~\cite{KraislerKronik13} does not account for the 5-fold degeneracy of d-levels. For the exact functional, when the homo is a p- or a d-orbital, changing the occupation numbers of the degenerate levels, keeping the number of electrons constant, does not change the energy. This can be viewed as analogous to piecewise linearity, but rather than changing the number of electrons, we change the occupations of the degenerate homo levels. Much like piecewise linearity, many common xc approximations do not satisfy this condition. The ensemble generalization is not expected to correct this deficiency, which therefore may contribute to the failure for d-levels but not s-levels.}

Certain atoms and ions were not considered in this work, as mentioned in the beginning of Sec.~\ref{sec:Results}. Their KS ground state is not a pure state: fractional occupations are involved, which means the KS ground state is an ensemble (although the interacting system may be in a pure state). Falling outside the scope of this work, these systems should definitely be approached in the future with a further generalized form of the present method.

From a broader perspective, the reformulation of the original results of Refs.~\cite{KraislerKronik13} and~\cite{KraislerKronik14} in Sec.~\ref{sec:Results:Eg_ions} (see particularly Figs.~\ref{fig:Eg_Li+} and~\ref{fig:Eg_O+}) provides a clear connection between four KS energy eigenvalues, which are, generally speaking, abstract mathematical quantities, and physical excitation energies. As the $\tilde{\eps}_\ho^\s$ relate to the IPs (of each spin channel), the quantities $a^\s$ relate to the EAs, using the appropriate shifts $w^\s$, which result from the ensemble treatment. One can even dare to further generalize this result to \emph{all} KS eigenvalues: all occupied levels $\eps_i$ shall be shifted by quantities similar to $v_0$ of Eq.~(\ref{eq:Hxc potential generalization}), where $\pphi_\ho$ is substituted by the relevant $\pphi_i$, and all vacant levels $\eps_a$ to be shifted by quantities similar to $w_0$ of Eq.~(\ref{eq:w_0_tau}), with a similar substitution. This suggestion requires proper theoretical justification and is the subject of our future work.

The improvement in the prediction of the IP and the fundamental gap demonstrated in this work on a large scale highlights the usefulness of the ensemble generalization of xc functionals in DFT, and reveals directions for further improvement. Due to the negligible numerical effort this method requires, it is expected to be useful and practically applicable also in molecular and nano-systems. 
%\red{In general, the approach developed here builds the bridge between the energy levels of the Kohn-Sham system -- generally, abstract mathematical objects -- and physically measured excitation energies, taking density functional theory to the next level, even with simple underlying xc approximations.}

\section{Conclusions}

In this work, we applied the ensemble generalization procedure~\cite{KraislerKronik13} to a variety of atoms and first ions to obtain the IP and the fundamental gap derived from quantities of the KS system. We used the LSDA and PBE as the underlying xc approximations, however our approach is applicable to \emph{any} xc approximation. As we demonstrated here, ensemble generalization improves the prediction of the IP and the fundamental gap. Notably, it also reintroduces the DD, even when it is absent from the underlying xc approximation.
We found our approach to be rather accurate when no d-orbitals are involved, and far less accurate otherwise. Development of further corrections to achieve further improvement is a subject of interest for future research.
%overcome this flaw is a subject of interest for future research.}

Due to the negligible numerical effort required by the ensemble correction method, it is expected to be useful and practically applicable also in molecules and nano-systems. The approach developed here builds the bridge between the quantities of Kohn-Sham DFT and physically measured excitation energies in finite systems, even with simple underlying xc approximations.

\section*{Supplemental Material}
The relevant data from our calculations is available in tabular form in the Supplemental Material. The tables include the IPs and fundamental gaps, as well as the homo and lumo energy levels, before and after the ensemble generalization.

\bibliography{bib_EK_2021_02_08}

\end{document}

\title{SUPPLEMENTARY MATERIAL for: \\
 “Ionization potentials and fundamental gaps in atomic systems from the ensemble-DFT approach”}

\author{Sharon Lavie}
%\email{Sharon.Lavie@mail.huji.ac.il}
\affiliation{Fritz Haber Center for Molecular Dynamics and Institute of Chemistry, The Hebrew University of Jerusalem, 9091401 Jerusalem, Israel}
\author{Yuli Goshen}
\affiliation{Fritz Haber Center for Molecular Dynamics and Institute of Chemistry, The Hebrew University of Jerusalem, 9091401 Jerusalem, Israel}
\author{Eli Kraisler}
\email{Author to whom correspondence should be addressed: eli.kraisler@mail.huji.ac.il}
\affiliation{Fritz Haber Center for Molecular Dynamics and Institute of Chemistry, The Hebrew University of Jerusalem, 9091401 Jerusalem, Israel}
\date{\today}

\maketitle

The four tables collected in this file include the numerical data that has been graphically presented and discussed in the main text.  
The tables refer to data for neutral atoms and first ions, calculated with the local spin-density approximation (LSDA) and with the Perdew-Burke-Ernzerhof Generalized Gradient Approximation (PBE-GGA).

Each table includes the ionization potential (IP), calculated from total energy differences, for each system; the total energy of each system; the quantum numbers $n$ and $l$ for the highest occupied (homo) energy level; the homo energy before and after the ensemble generalization, as discussed in the main text, and similarly for the lowest unoccupied (lumo) level. The aforementioned information is provided for both the up- and the down-spin channels.
Tables for ions include also the fundamental gap, $E_g$, calculated from total energy differences.
In addition, the tables include experimental data for the IP and $E_g$, for comparison.

All energy values are given in atomic Hartree units.

%\newpage

%.

\newpage

\includepdf[pages=-]{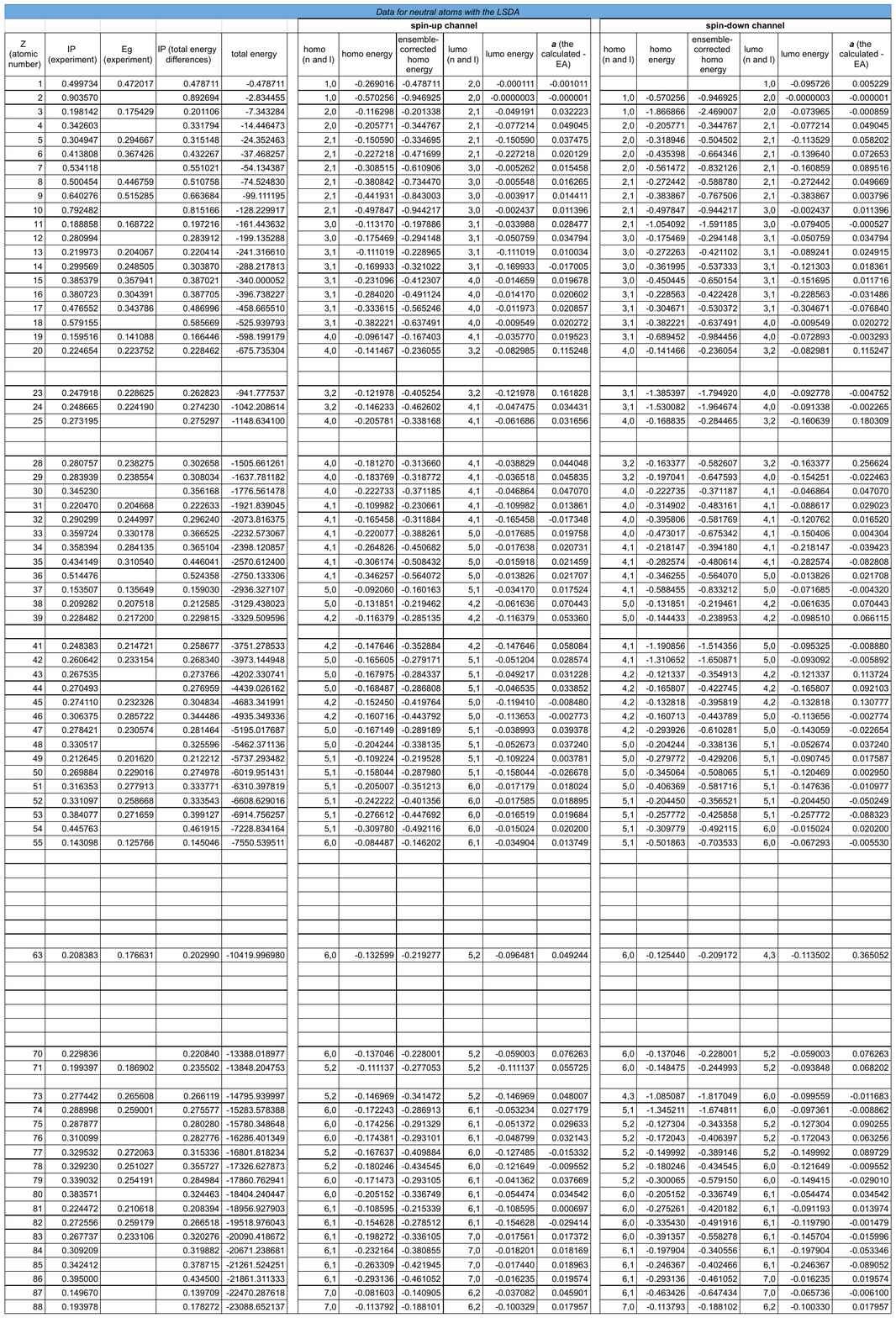}

\includepdf[pages=-]{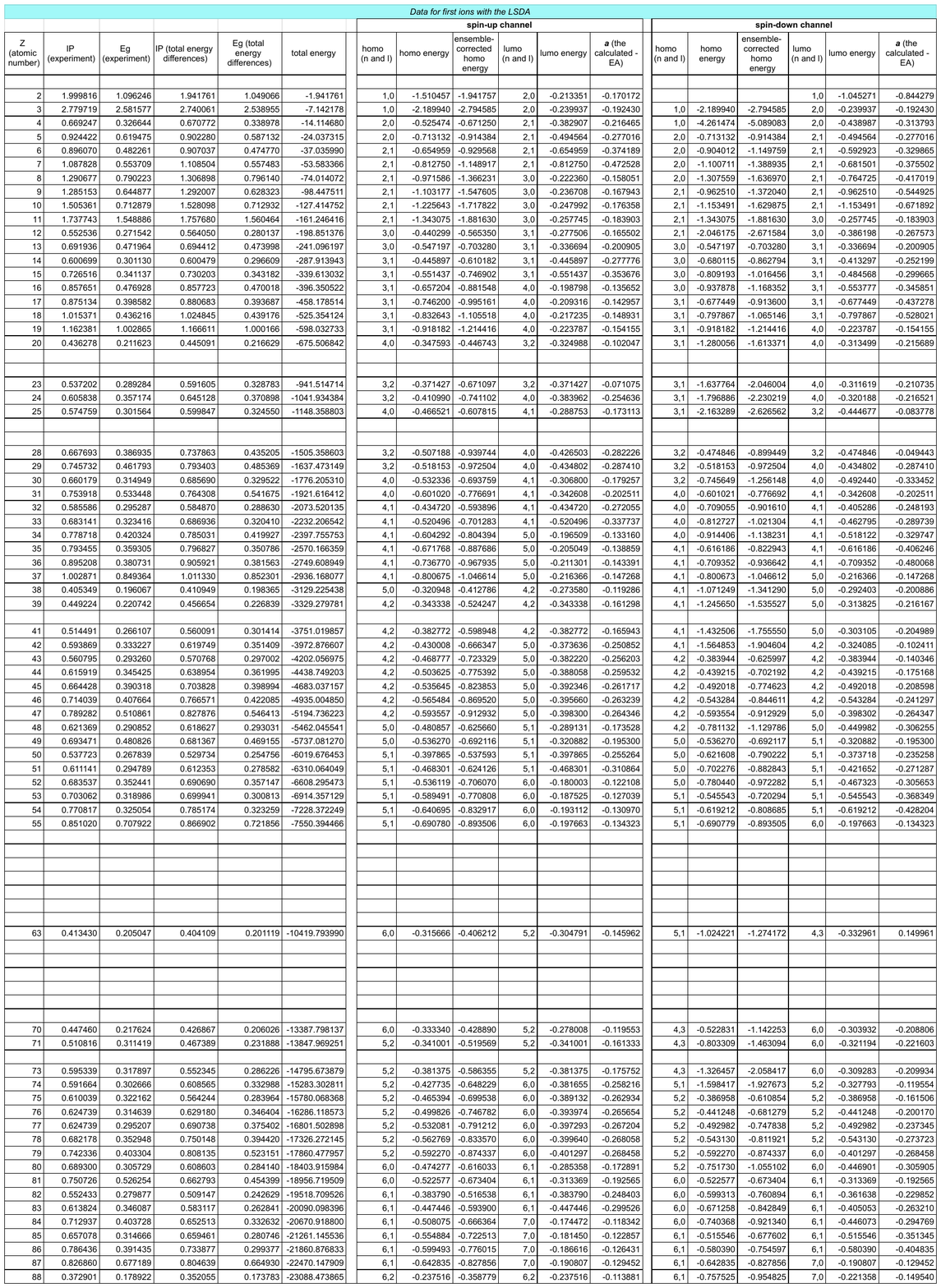}

\includepdf[pages=-]{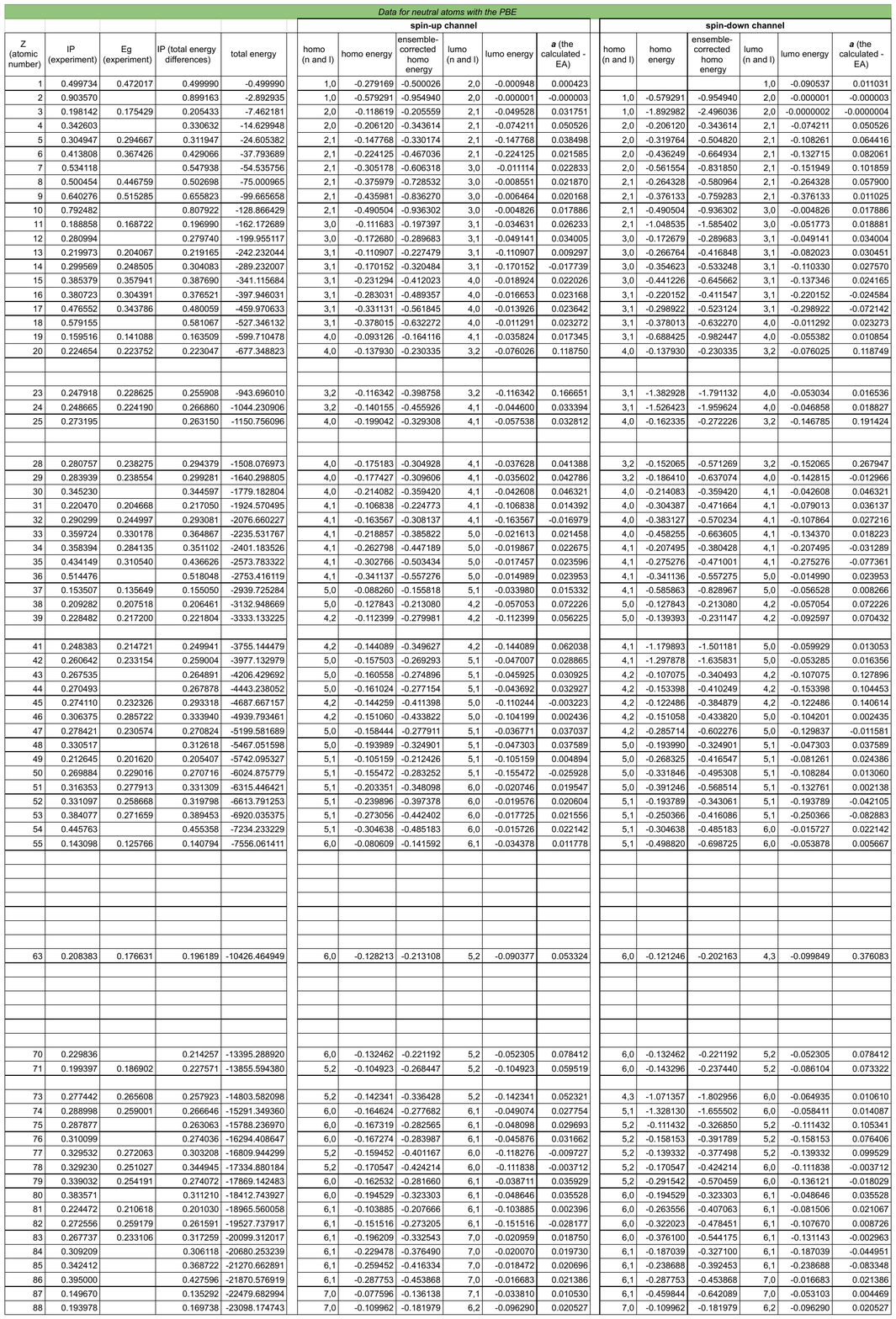}

\includepdf[pages=-]{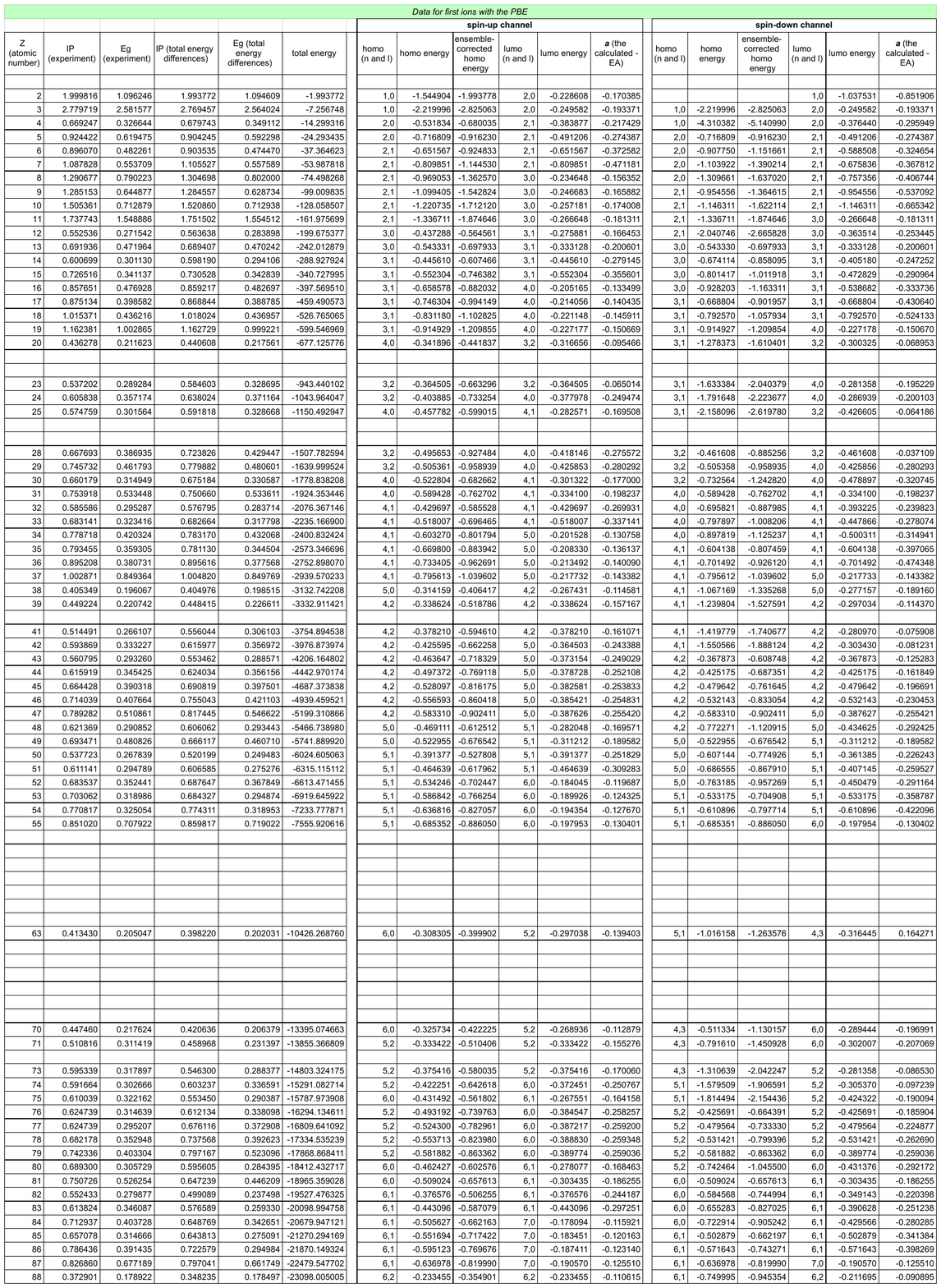}

%\bibliography{bib_EK_2021_02_08}